\documentclass[11pt,english]{article}
\usepackage[T1]{fontenc}
\usepackage[latin9]{inputenc}
\usepackage[letterpaper]{geometry}
\usepackage{color}
\usepackage{float}
\usepackage{textcomp}
\usepackage{amstext}
\usepackage{graphicx}
\usepackage{amssymb}
\usepackage{esint}
\usepackage[authoryear]{natbib}

\makeatletter

\providecommand{\tabularnewline}{\\}

\date{}

\makeatother

\usepackage{babel}

\begin{document}
\bibliographystyle{plainnat}

\pagestyle{empty}

\centerline{\includegraphics{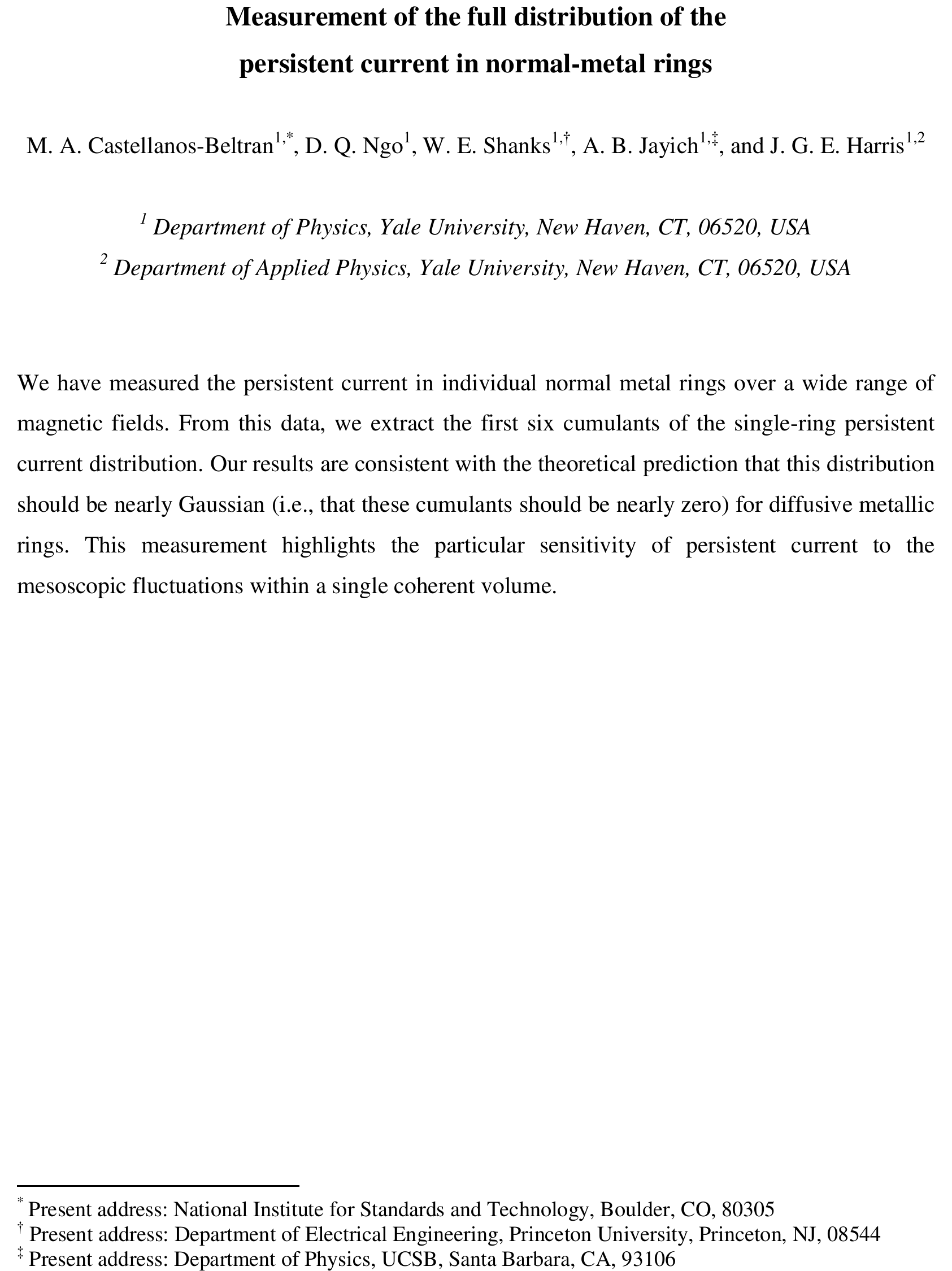}}
\newpage

\advance\voffset by -1.5cm

\centerline{\includegraphics{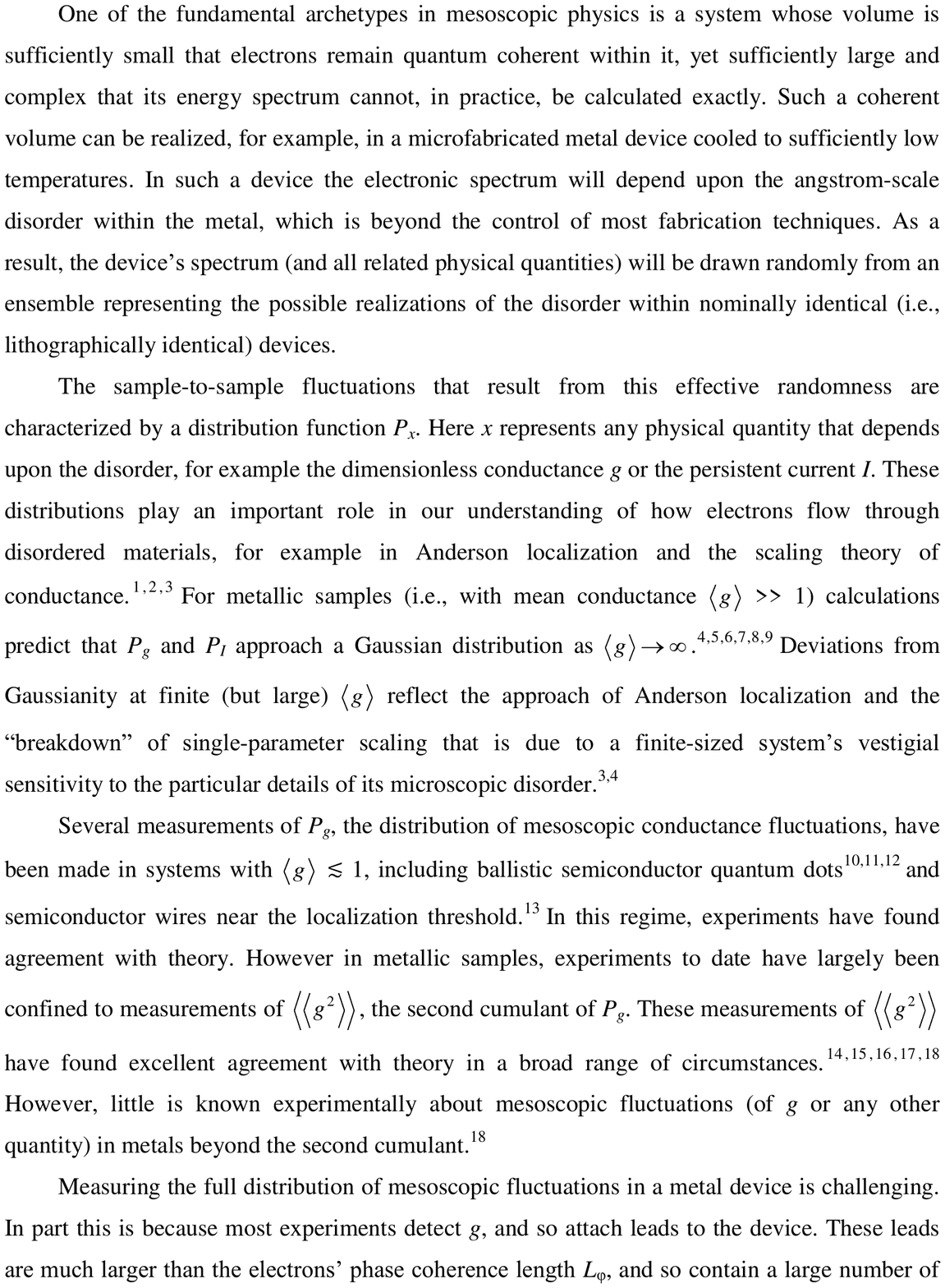}}
\newpage

\centerline{\includegraphics{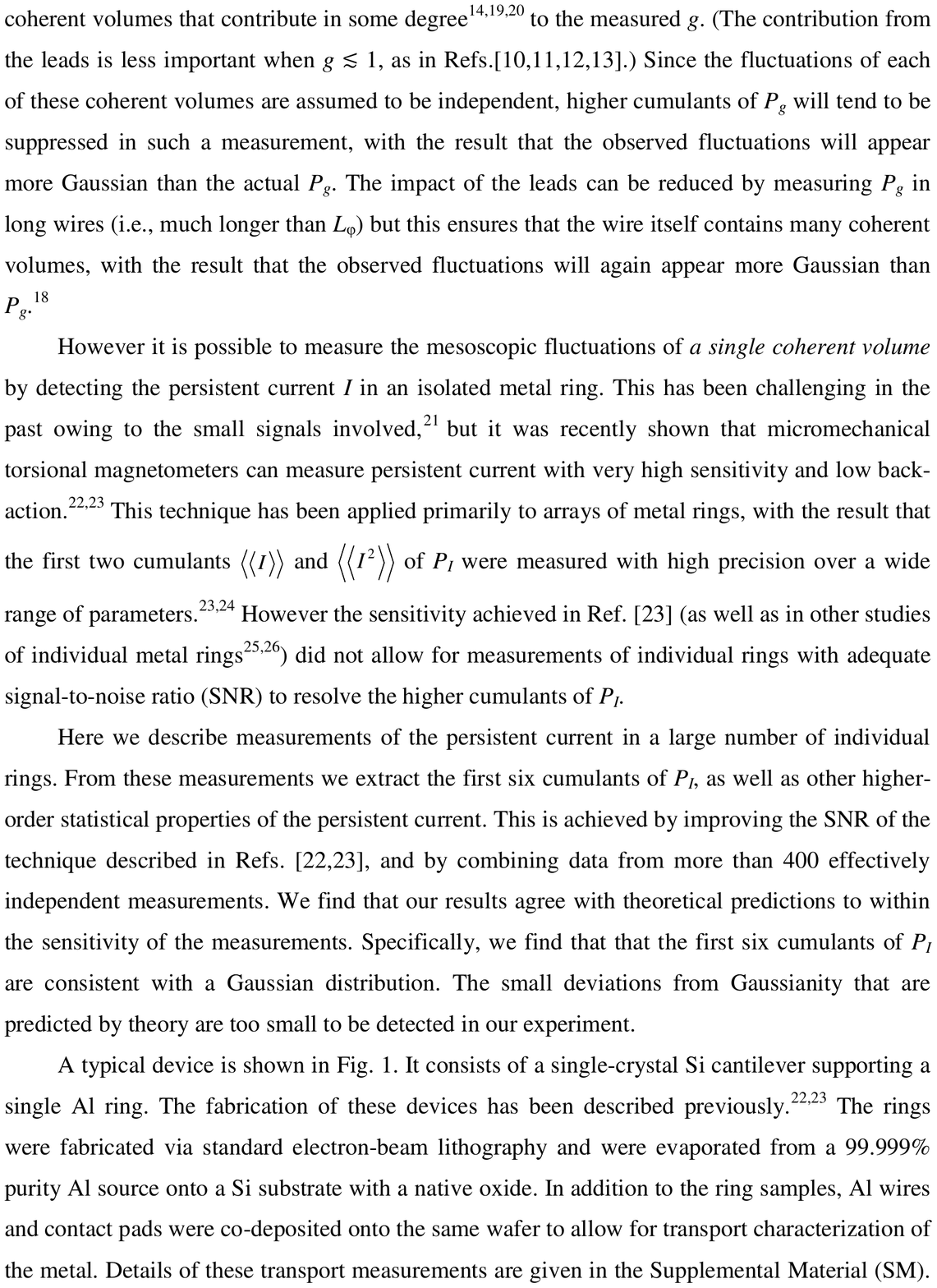}}
\newpage

\centerline{\includegraphics{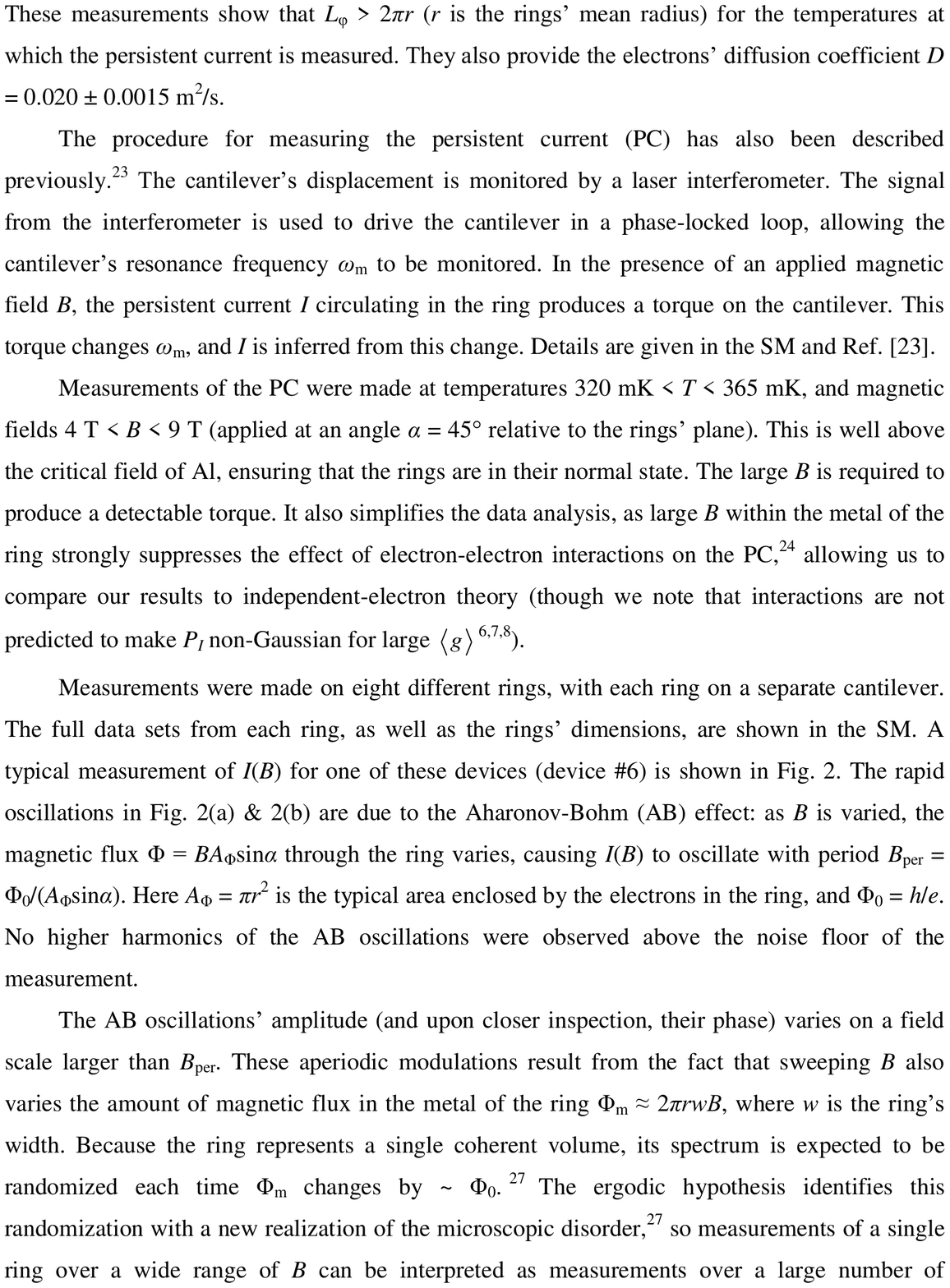}}
\newpage
\centerline{\includegraphics{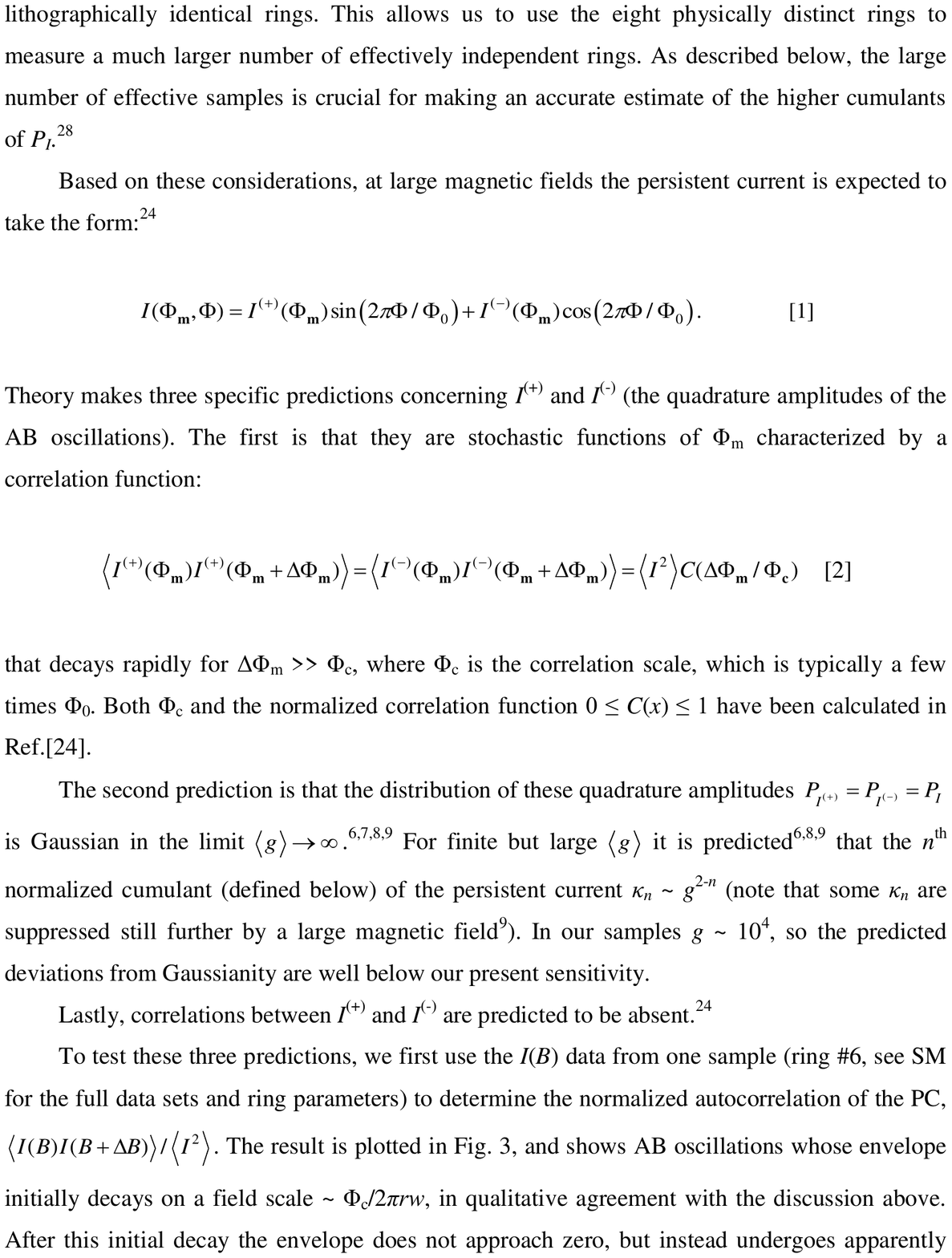}}
\newpage
\centerline{\includegraphics{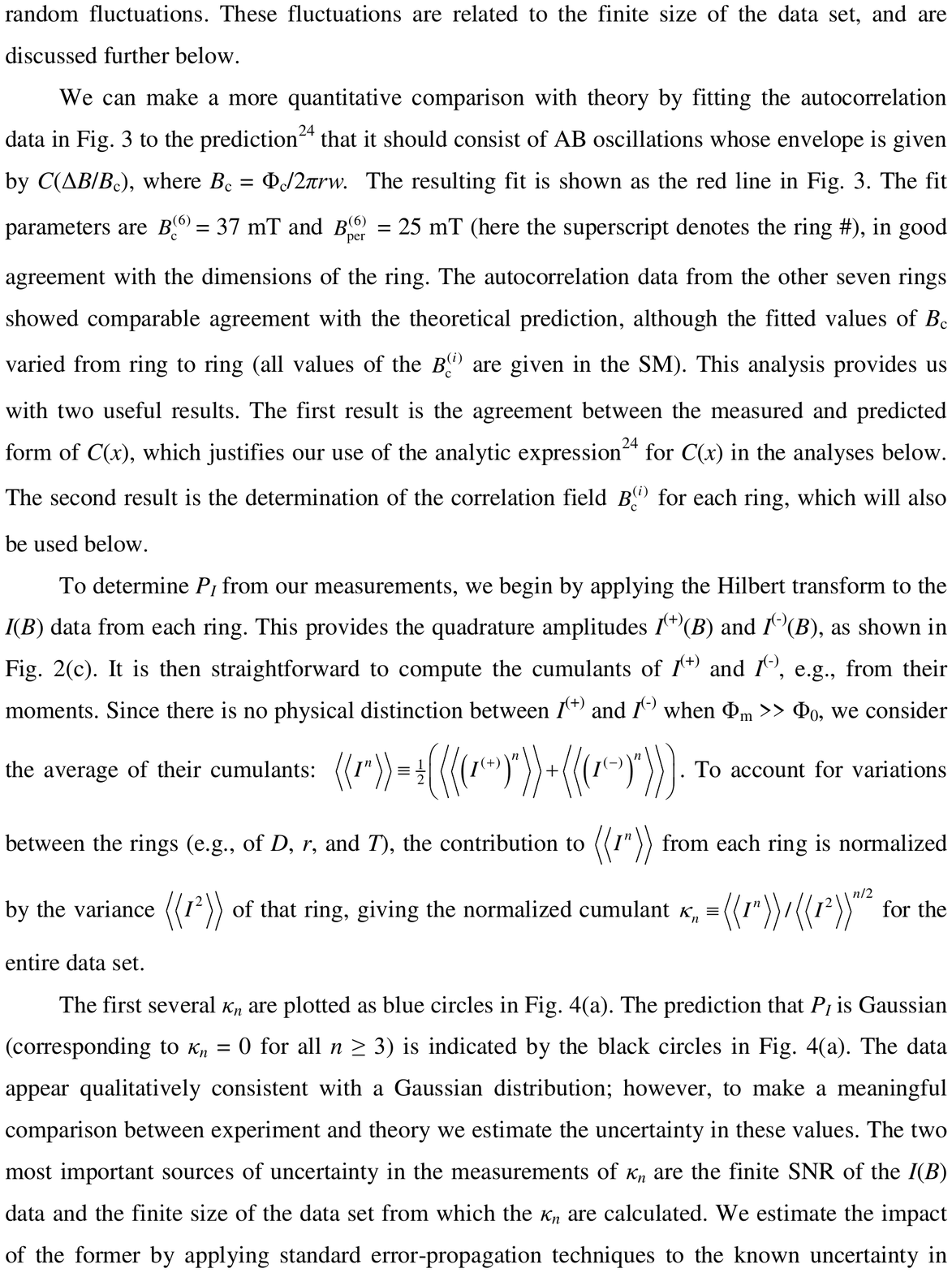}}
\newpage
\centerline{\includegraphics{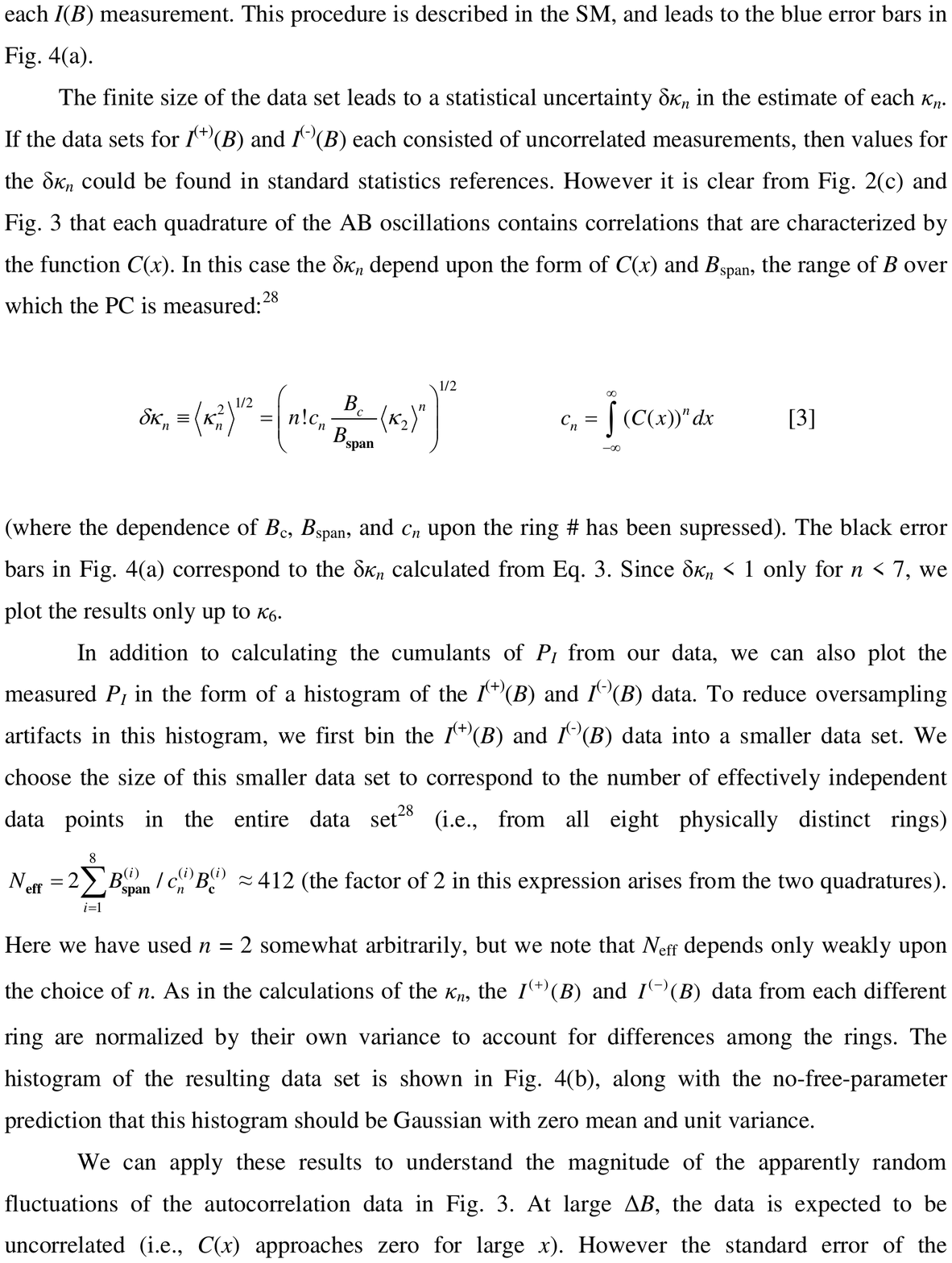}}
\newpage
\centerline{\includegraphics{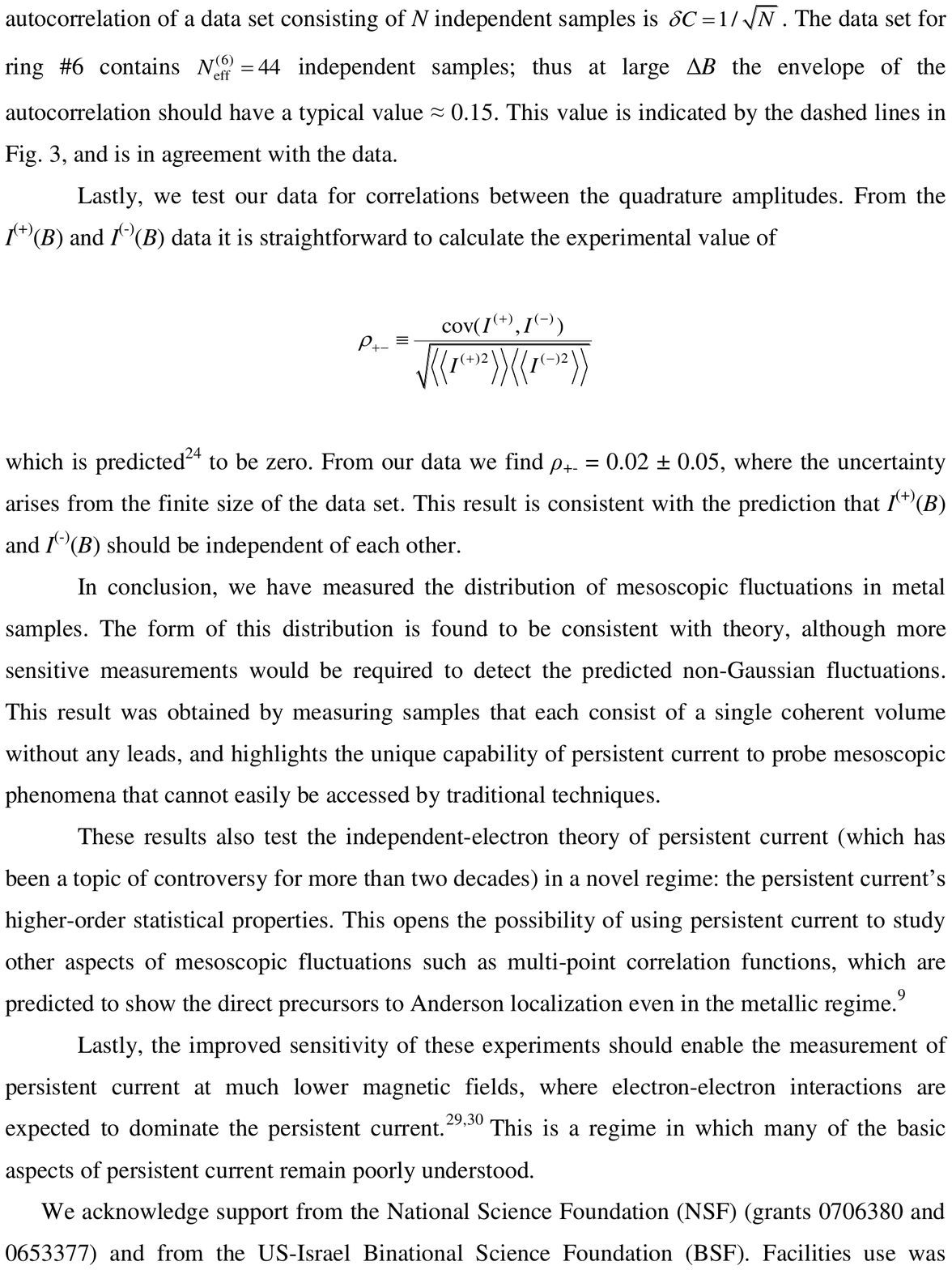}}
\newpage
\centerline{\includegraphics{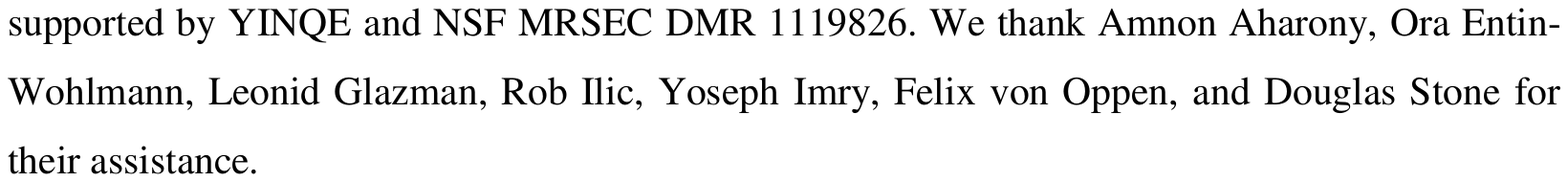}}
\newpage
\centerline{\includegraphics{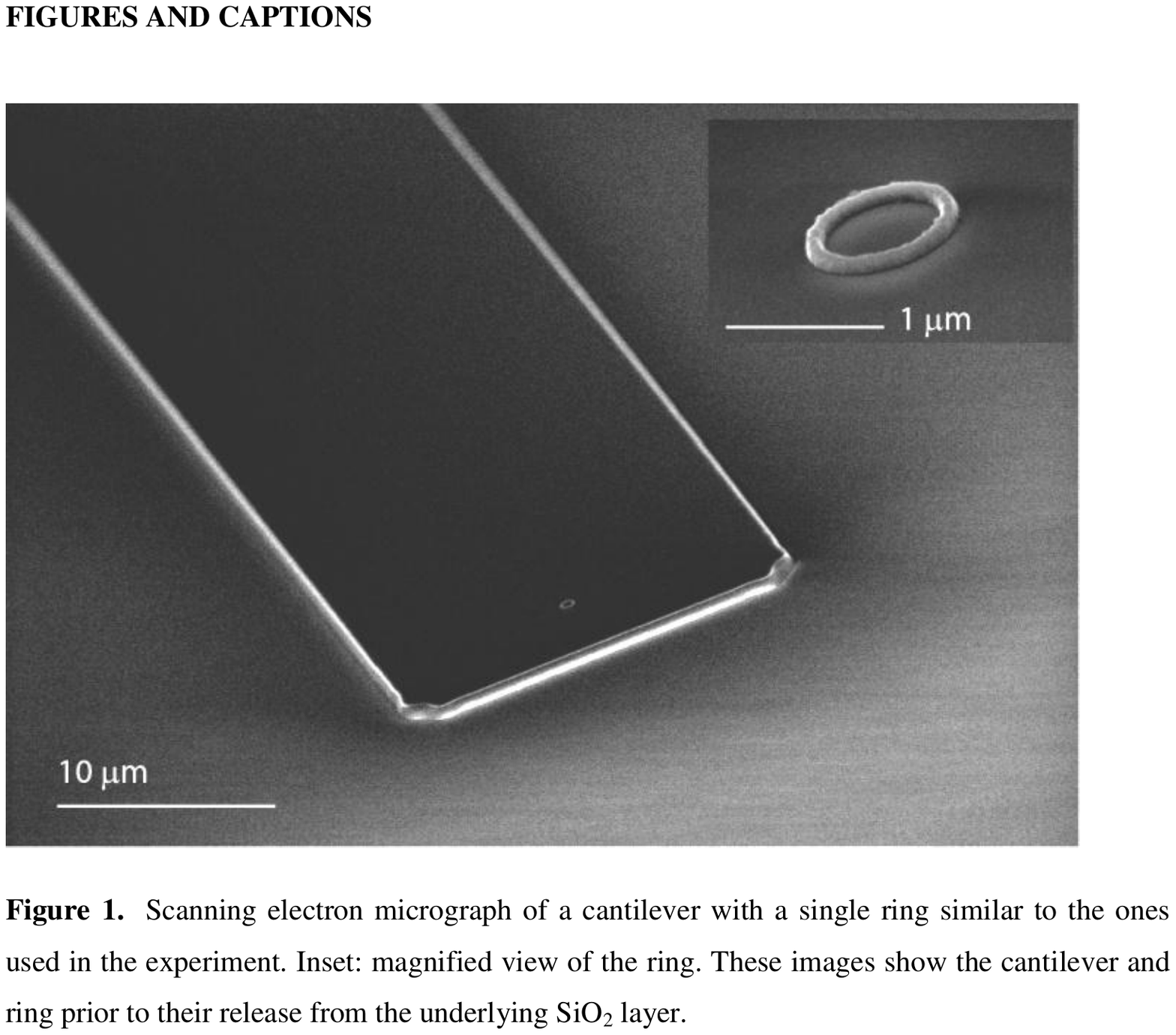}}
\newpage
\centerline{\includegraphics{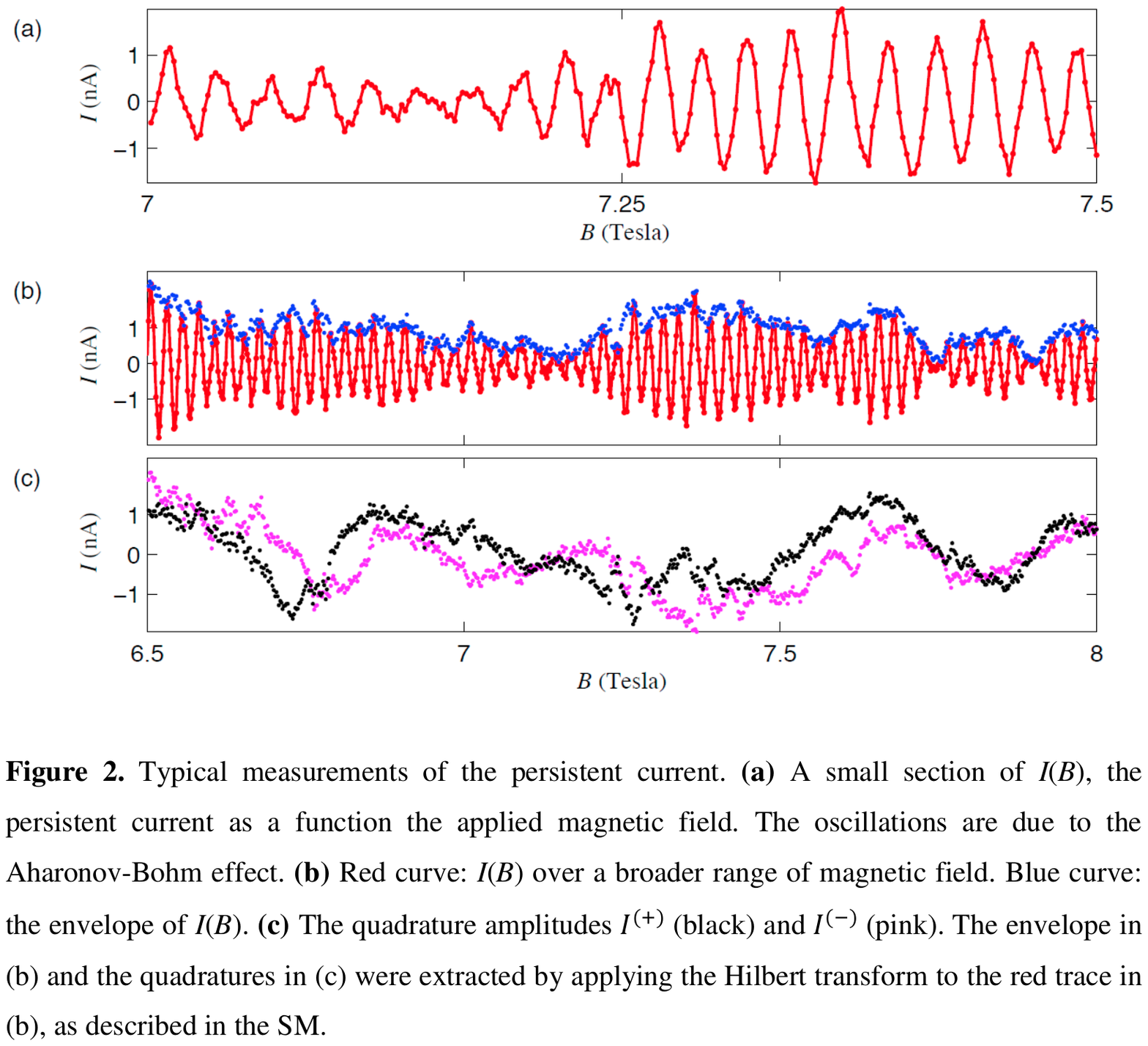}}
\newpage
\centerline{\includegraphics{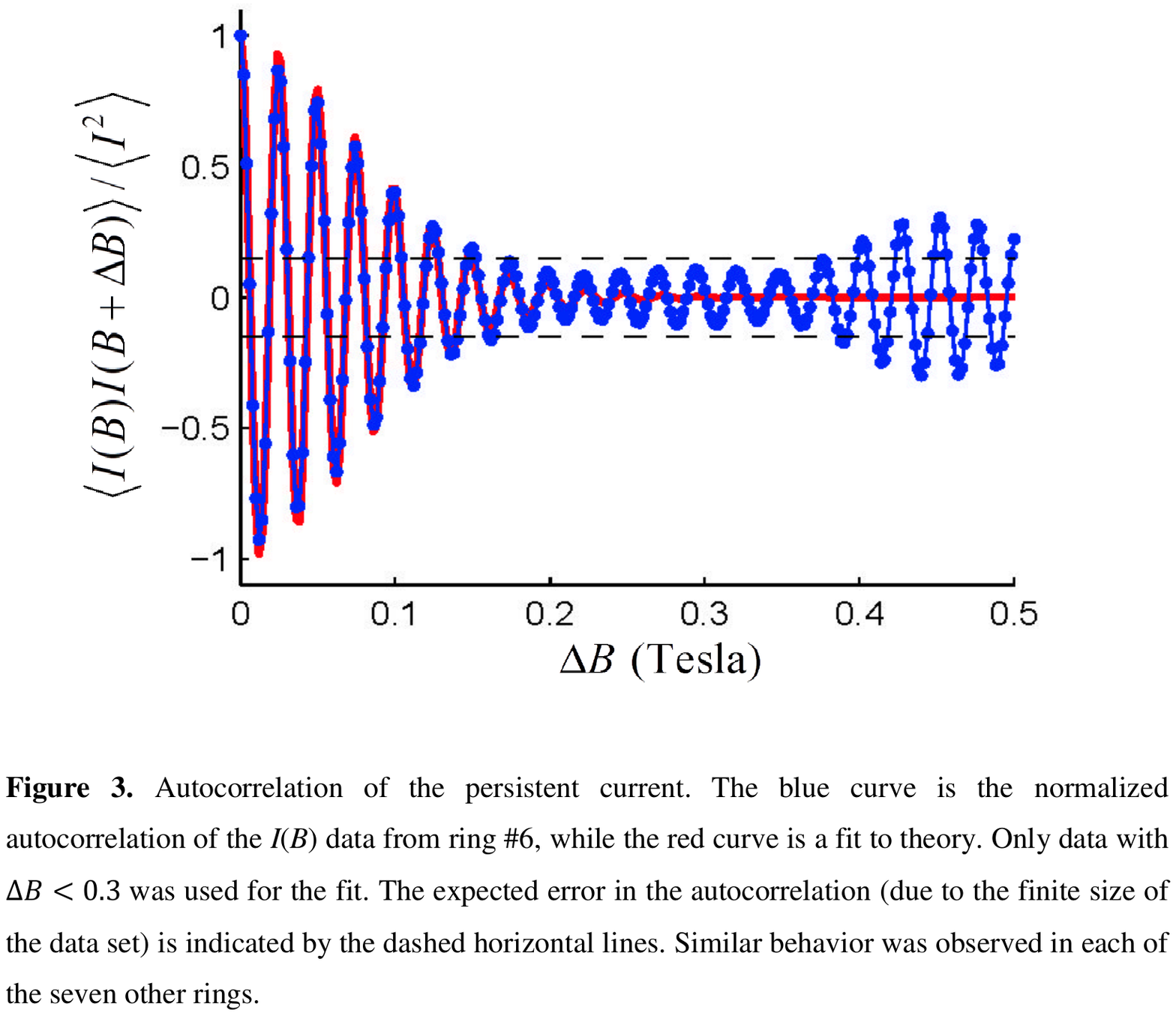}}
\newpage
\centerline{\includegraphics{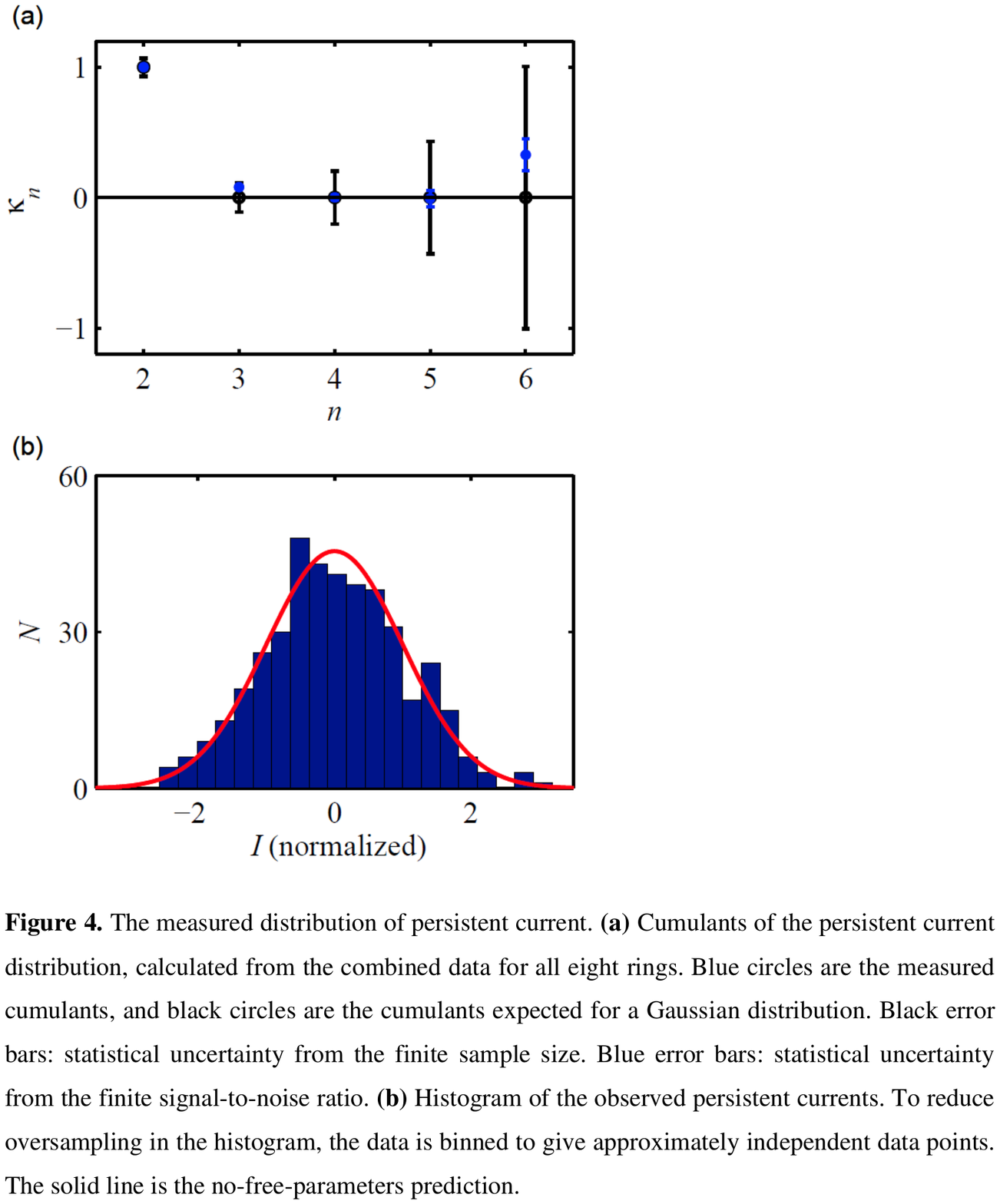}}
\newpage
\centerline{\includegraphics{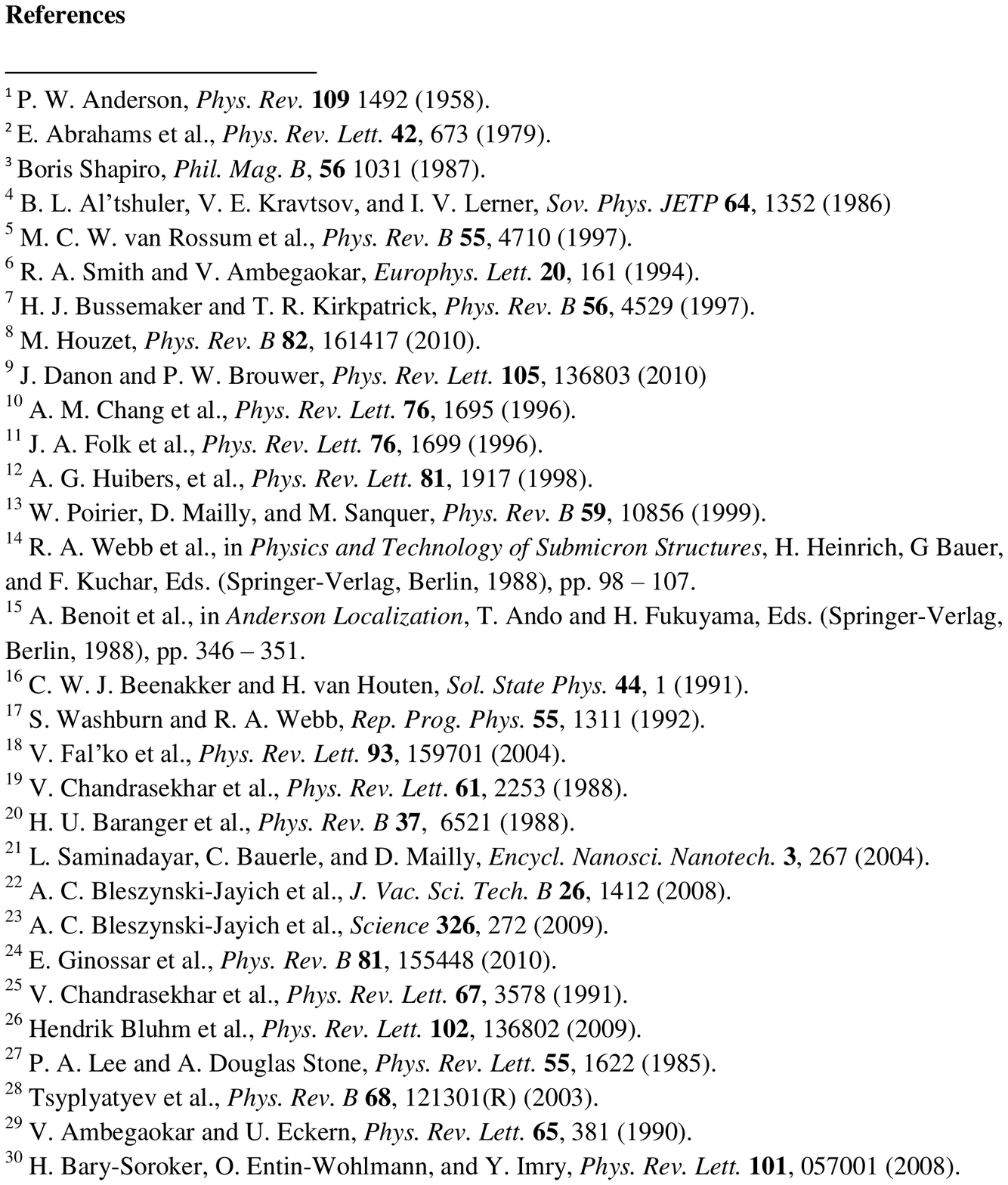}}
\newpage

\advance\voffset by 1.5cm

\pagebreak{}

\pagestyle{plain}
\setcounter{page}{1}
\pagenumbering{arabic}

\title{Measurement of the full distribution of the persistent current in
normal-metal rings: Supplementary Information.}

\author{M. A. Castellanos-Beltran,$^{\text{1}}$%
\thanks{To whom correspondence should be addressed; E-mail:  manuel.castellanosbeltran@nist.gov.%
}~~D. Q. Ngo,$^{\text{1}}$ ~W. E. Shanks,$^{\text{1}}$~A. B.
Jayich,$^{\text{1}}$\\
and J. G. E. Harris$^{\text{1,2}}${\small }\\
{\small $^{\text{1}}$Department of Physics, Yale University,
New Haven, CT, 06520, USA }\\
{\small $^{2}$Department of Applied Physics, Yale University,
New Haven, CT, 06520, USA}}

\maketitle

\section{Transport measurements}

The main paper compares our measurements of the persistent current
in normal metal rings to predictions of a theoretical model based
on diffusive, non-interacting electrons\citet{Ginossar2010}. This
theory depends explicitly upon two parameters of the rings: their
diffusion coefficient $D$ and their spin-orbit scattering length
$L_{SO}$. This theory also implictly assumes that the rings' phase
coherence length ($L_{\phi}$) is much greater than the rings' circumference
$L$. The persistent current data presented in the main paper can
be compared with this theory by taking $D$ and $L_{SO}$ as fitting
parameters (and assuming $L_{\phi}\gg L$), but in order to constrain
this comparison more tightly, we have used transport measurements
to directly measure $D$, $L_{SO}$, and $L_{\phi}$. Transport measurements
similar to those described in Ref.\citet{Ania_Science} were performed
on an aluminum wire codeposited with the rings studied in this article.
The wire was deposited on the same wafer as the rings. It had a length
$L=255\:\mathrm{\mu m}$, a width $w=105\:\mathrm{nm}$ (determined
from SEM images) and a thickness $t=115\:\mathrm{nm}$ (determined
from AFM measurements).

\begin{figure}
\begin{centering}
\includegraphics{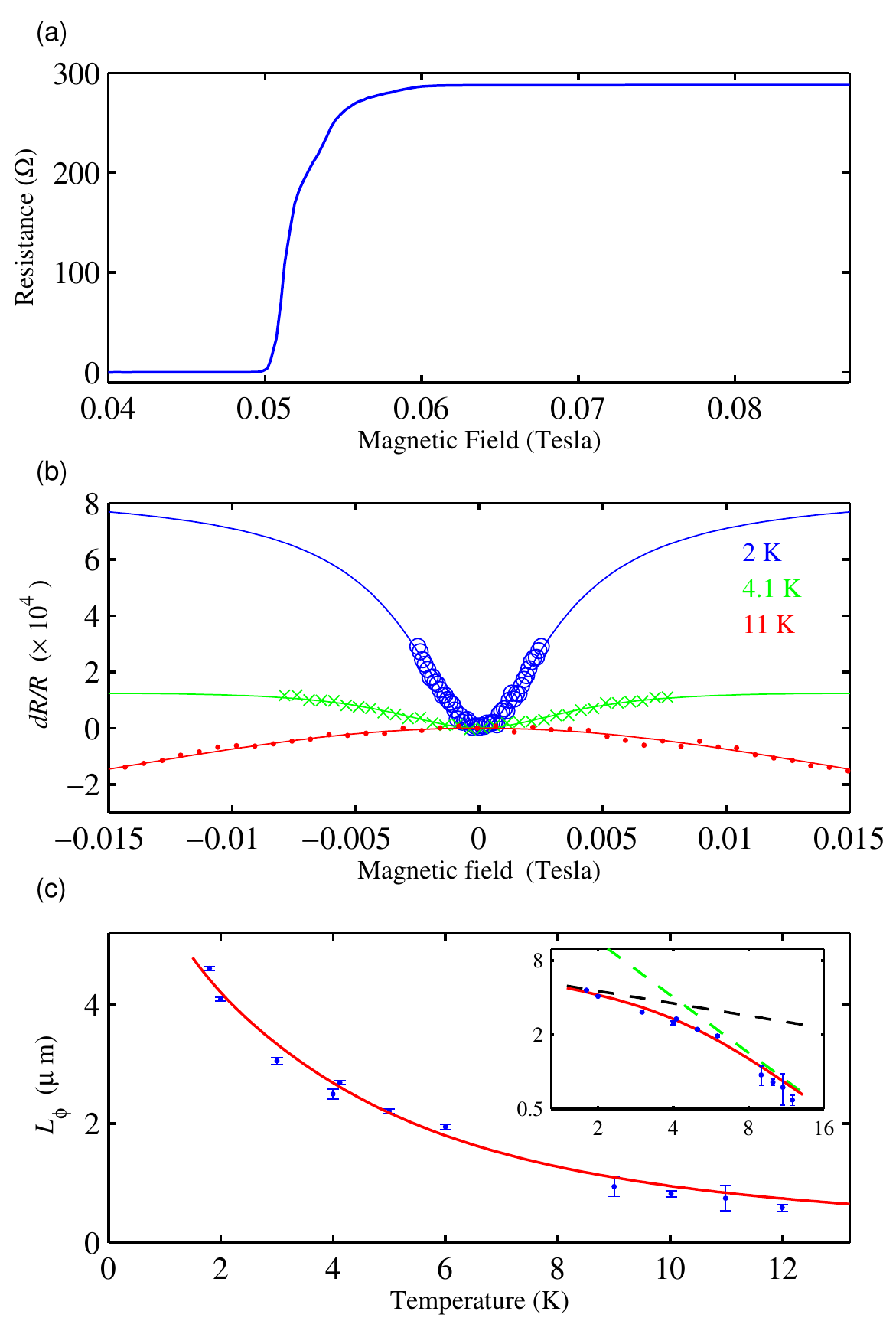}
\par\end{centering}

\centering{}\caption{\label{fig:Transmeas}(a) Resistance measurement of the aluminum wire
as the magnetic field is swept through the critical field $B_{c}$
at a temperature $T=300\,\mathrm{mK}$. The measured resistance in
the normal state was 288 $\mathrm{\Omega}$. Due to the loss of one
the leads, in series with the sample, we measured a lead resistance
of 22 $\mathrm{\Omega}$. This finite resistance is observed even
when at $|B|<B_{c}$ and $T<T_{c}$, and it has been subtracted from
the trace. (b) Magnetoresistance data for $T=2\;\mathrm{K}\:(\circ)$,
$4.1\;\mathrm{K}\:(\times)$ and $11\;\mathrm{K}\:(.)$ and fits (solid
lines). The trace for $\mathrm{11\: K}$ has been multiplied by 20.
At low temperatures, superconducting fluctuations are the dominant
effect observed in the magnetoresistance measurements. The weak localization
becomes the dominant effect at $T>8\:\mathrm{K}$. (c) Phase coherence
length $L_{\phi}$ extracted from fitting the magnetoresistance data
as a function of temperature (points) and fit (solid line) to the
expected coherence length assuming both electron-electron and electron-phonon
contributions. Inset: Log plot showing $\tau_{e-p}$ and $\tau_{ee}$
contribution as dashed-green and dashed-black lines respectively. }

\end{figure}

Measurements of the wire's resistance were performed using a bridge
circuit similar to that of Ref.\citet{Chandrasekhar}: a three terminal
arrangement was used here as one of the four original leads was unintentionally
blown out as the sample cooled to 4.2 K. One side of the sample was
connected to the bridge and the voltage probe through separate leads,
while the other side of the sample was connected to ground and the
bridge through the same lead. The sample resistance was distinguished
from the lead resistance by measuring the resistance of the sample
as it transitioned from the superconducting to the normal state as
the magnetic field was swept at a temperature below the wire's transition
temperature $T_{c}.$ A measurement of this transition is shown in
Fig. \ref{fig:Transmeas}a.

The diffusion constant $D$ was then calculated using the Einstein
relation ($\rho^{-1}=e^{2}gD$), with $e$ being the electron charge
and $g$ the electron density of states per unit volume at the Fermi
level for aluminum\citet{Ashcroft}. The resistance was measured to
be 288 $\mathrm{\Omega}$. Based on the film dimensions, we infer
a resistivity $\rho=1.36\times10^{-8}\pm0.1\:\mathrm{\Omega m}$.
This corresponds to a diffusion constant $D_{\rho}=0.02\pm0.0015\:\mathrm{m^{2}/s}$.
We label $D_{\rho}$ the diffusion constant extracted from this measurement
of $\rho$.

We also performed magnetoresistance measurements on the same wire
at temperatures above $T_{c}$ in order to extract $L_{\phi}$ and
$L_{SO}$. The magnetoresistance in the aluminum wire has contributions
from weak localization as well as Maki-Thompson superconducting fluctuations.
The analytical form for the weak-localization correction to the resistance
of a 1-D wire in a magnetic field is given by\begin{equation}
\frac{dR^{WL}}{R}=\frac{R(B)-R(B=0)}{R(B=0)}=\frac{3}{2}f_{1}\left(B,b(L_{2})\right)-\frac{1}{2}f_{1}\left(B,b(L_{\phi})\right)\label{eq:1}\end{equation}
where $1/L_{2}^{2}=1/L_{\phi}^{2}+4/3L_{SO}^{2}$, $L_{\phi}$ is
the electron phase coherence and $L_{SO}$ is the spin-orbit coupling
length. The functions $f_{1}(B,B1)$ and $b(l)$ are defined in Ref.\citet{Ania_Science}.
The Maki-Thompson contibrution is given by \begin{equation}
\frac{dR^{MT}}{R}=-\beta\left(\frac{T}{T_{c}}\right)f_{1}\left(B,b(L_{\phi})\right)\label{eq:2}\end{equation}
where $\beta(t)$ is a function that diverges logarithmically when
$t\rightarrow1$ (see Ref.\citet{Ania_Science} and references therein).
Equation \ref{eq:2} is only valid provided \begin{equation}
B\ll\frac{k_{B}T}{4De}\ln\left(T/T_{c}\right).\label{eq:2b}\end{equation}
Fits were done using the sum of both equations \ref{eq:1} and \ref{eq:2}:\begin{equation}
\frac{dR}{R}=\frac{dR^{WL}}{R}+\frac{dR^{MT}}{R}\label{eq:dRtotal}\end{equation}

Magnetoresistance measurements were made at a series of temperatures
above $T_{c}$ between 1.8 and 12 K. $L_{SO}$ was obtained from fits
of the magnetoresistance to Eq. \ref{eq:dRtotal} measured at the
highest temperatures ($T=9-12\:\mathrm{K}$), where $L_{SO}$ was
measured to be $1.54\pm0.06\:\mathrm{\mu m}$.%
\footnote{\textcolor{black}{The error bar in $L_{SO}$ was obtained from the
following analysis: each of the fits to the four traces at the tempratures
$T=9$, 10, 11 and 12 K provided the same $L_{SO}$ within the fit
error. The errorbars from the obtained values of $L_{SO}$ at each
temperature are based on the goodness of the fits. Then we do a weighted
average of the obtained values of $L_{SO}$ and the error is correspondingly
$\sigma=1/\sqrt{\Sigma_{i=1}^{4}1/\sigma_{i,fit}^{2}}$ .}%
} For fits of the lower temperature data, $L_{SO}$ was fixed at this
value, so $L_{\phi}$ was the only fitting parameter. Fits to the
magnetoresistace data using Eq. \ref{eq:dRtotal} are shown in Fig.
\ref{fig:Transmeas}b. The constraint set by Eq. \ref{eq:2b} upon
the validity of Eq. \ref{eq:2} sets the limit over which the fits
were performed.

The temperature dependence of $L_{\phi}$ can be explained based on
the different contributions to the inelastic collisions of the electrons.
Theory predicts that $\tau_{\phi}$ is limited by inelastic collisions
with other electrons through the screened Coulomb interactions ($\tau_{ee}$),
with phonons ($\tau_{ep}$) and with extrinsic sources such as magnetic
impurites. The latter should be negligible for the high purity aluminum
source used and since no magnetic impurity has been observed to behave
as a localized moment when dissolved in aluminum\citet{Ashcroft}.
The fitted values of $L_{\phi}$ as a function of temperature are
shown in Fig. \ref{fig:Transmeas}c. The function used to fit to the
data is $L_{\phi}=\sqrt{D\tau_{\phi}}$, where \begin{equation}
\tau_{\phi}^{-1}=A_{ep}T^{3}+A_{ee}T^{2/3}.\label{eq:phasetime}\end{equation}
$A_{ep}$ and $A_{ee}$ are fit parameters. The first term of Eq.
\ref{eq:phasetime} corresponds to the electron-phonon scattering
rate, $\tau_{ep}^{-1}=A_{ep}T^{3}$. From our fit we find that $A_{ep}=2.0\pm0.2\times10^{7}\mathrm{s^{-1}K^{-3}}$
which is within a factor of two of previoulsly measured electron-phonon
coefficients for comparable Aluminum wires\citet{Ania_Science,Santhanam}.

The second term of Eq. \ref{eq:phasetime} corresponds to electron-electron
scattering. The wire (and rings) studied here had a width and thickness
smaller than $L_{\phi}$. Therefore, the quasi-1D prediction for electron-electron
interaction applies and the expression for $\tau_{ee}$ according
to Ref.\citet{Birge} is \begin{equation}
\tau_{ee}^{-1}=A_{ee}T^{2/3}=\left[\frac{R_{\square}e^{2}k_{B}\sqrt{D}}{2\sqrt{2}\hbar^{2}w}\right]^{2/3}T^{2/3}.\label{eq:3}\end{equation}
The fitted valued for $A_{ee}$ is $0.65\pm0.03\times10^{9}\mathrm{s^{-1}K^{-2/3}}$.
The expected theoretical value based on Eq. \ref{eq:3} and a diffusion
constant of $D=0.020\:\mathrm{m^{2}/s}$ is $A_{ee}=0.15\times10^{9}\mathrm{s^{-1}K^{-2/3}}$.
Although there is a large discrepancy betwen the expected and fitted
value of $A_{ee}$, the primary purpose of these transport measurements
is to show that $L_{\phi}$ is greater than the circumference of the
rings $L=1.8$ and 2.8 $\mathrm{\mu m}$; therefore we did not look
further into this disagreement.%
\footnote{ In addition, it has been pointed out that close to the superconducting
transition, the electron-electron inelastic scattering rate can be
modified due to superconducting fluctuations\citet{Santhanam}. These
fluctuations may alter the numerical coefficient $A_{ee}$.%
}

\begin{figure}
\begin{centering}
\includegraphics{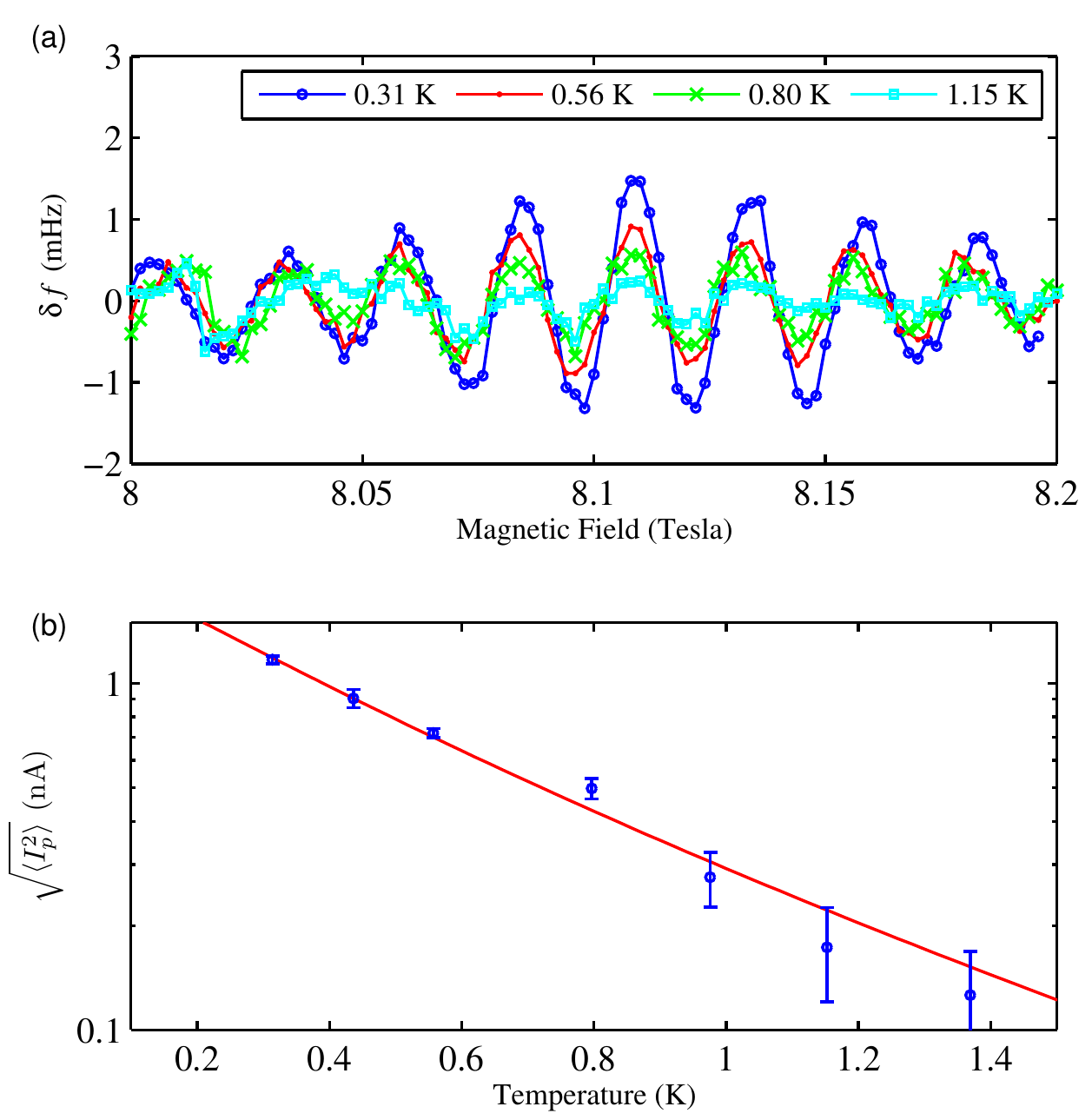}
\par\end{centering}

\centering{}\caption{\label{fig:FrequencyshiftTscan}(a) Frequency shift versus magnetic
field for a ring of radius $r=296$ nm at different temperatures.
The amplitude of frequency oscillations decreases with increasing
temperature as expected. The amplitude of the frequency oscillations
is converterted into current oscillations amplitude and plotted as
a function of temperature in (b). Data (circles) and a fit (solid
line) based on analysis similar to that explained in Ref.\citet{Ginossar2010}
are plotted. From the fit we estimate a diffusion constant of $D=0.021\pm0.02$. }

\end{figure}

The value for the diffusion constant extracted from the resistivity
measurement can then be compared with the diffusion constant extracted
from the persistent current measurement (Fig. \ref{fig:FrequencyshiftTscan}).
Using a similar analysis to that explained in Ref. \citet{Ginossar2010}
we extract a value of $D_{PC}=0.021\pm0.002\:\mathrm{m^{2}/s}$. We
label this value as $D_{PC}$ to distinguish from $D_{\rho}$. $D_{PC}$
and $D_{\rho}$ differ by about $5\%$. This seems reasonable given
differences between wire cross section and ring cross section plus
statistical uncertainty in $D_{PC}$.

\section{Data Analysis}

We inferred the persistent current in each ring by measuring the resonance
frequency of the cantilever into which the ring was integrated by
driving the cantilever in a phase-locked loop\citet{Albrecht}. The
method used to infer the current is explained in the supplementary
information of Ref.\citet{Ania_Science}. Here we briefly review this
analysis.

As it was shown in Ref.\citet{Ginossar2010}, the presence of magnetic
field $B_{m}$ inside the metal of the ring leads to an aperiodic
modulation of the persistent current oscillations and a change in
the fl{}ux dependence of of the persistent current from the simpler
case where there is only a pure Aharonov-Bohm flux $\phi$ threading
the ring. This change is the modification of the time-reversal relation
from $I(\phi)=-I(-\phi)$ to $I(B_{m},\phi)=-I(-B_{m},-\phi)$ . As
a result, the current is no longer odd in the Aharonov-Bohm fl{}ux
$\phi$ and it takes the more general form at fi{}xed $B_{m}$ \begin{equation}
I(\phi)=\sum_{p}I_{h/pe}^{+}\sin(2\pi p\frac{\phi}{\Phi_{0}})+I_{h/pe}^{-}\cos(2\pi p\frac{\phi}{\Phi_{0}})\label{eq:Ifourierseries}\end{equation}
where the variables $I_{h/pe}^{+}$ and $I_{h/pe}^{-}$ are stochastic
variables that vary with the magnetic field $B_{m}$ as well as with
microscopic disorder, and $\Phi_{0}=h/e$ is a flux quantum. Determining
the distribution of $I_{h/pe}^{+}$ and $I_{h/pe}^{-}$ is the central
point of the main paper.

We monitor the the cantilever frequency as we vary the magnetic field.
The cantilever's deflection leads to a rotation of the sample, responsible
for the coupling between the persistent current and the cantilever.
When $\theta\neq\pi/2$ (where $\theta$ is the angle between the
plane of the ring and the applied magnetic field $B$), the frequency
change is dominated by the following term: \begin{eqnarray}
\delta f & = & -\frac{ABf_{0}\gamma_{m}}{kLx_{\mathrm{tip}}}\cos(\theta)\sum_{p}J_{1}\left(2\pi p\frac{AB\cos(\theta)}{\Phi_{0}}\frac{\gamma_{m}x_{\mathrm{tip}}}{L}\right)\times\label{eq:deltaf}\\
 &  & \left(I_{h/pe}^{+}\cos(2\pi p\frac{BA\sin\theta}{\Phi_{0}})-I_{h/pe}^{-}\sin(2\pi p\frac{BA\sin\theta}{\Phi_{0}})\right)\nonumber \end{eqnarray}
where $A$ is the area of the ring, $f_{0}$ is the natural resonance
frequency of the cantilever, $\gamma_{m}$ is the ratio between the
slope of the cantilever and the factor $x_{\mathrm{tip}}/L$ for the
flexural mode $m$. For $m=\{1,2\}$, $\gamma_{m}=\{1.377,4.788\}$.
$k$ is the cantilever spring constant, $L$ is the length of the
cantilever, $x_{\mathrm{tip}}$ is the amplitude of oscillation at
the tip of the cantilever, and $J_{1}(x)$ is the first Bessel function
of the first kind.

At the temperatures of our experiment the current is dominated by
the first harmonic, $p=1$. Thus, for the analysis of the data shown
in the main paper we use what in the supplementary online information
(SOI) of Ref.\citet{Ania_Science} is refered to as method B: we assume
that the signal only has the $p=1$ component in Eq. \ref{eq:deltaf}
and that the argument of the Bessel function varies only weakly over
a given data set. In that case, the change in frequency is essentially
the derivative of the persistent current (again ignoring all terms
for $p>1$). \[
\frac{\partial I}{\partial B}\thickapprox-\delta f\frac{2\pi A\sin\theta}{\Phi_{0}}\left[\frac{AB\nu\gamma_{m}}{kLx_{\mathrm{tip}}}\cos(\theta)J_{1}\left(2\pi p\frac{AB\cos(\theta)}{\Phi_{0}}\frac{\gamma_{m}x_{\mathrm{tip}}}{L}\right)\right]^{-1}\]
In order to estimate the current this quantity can be numerically
integrated. However, since we are interested in the statisitics of
the variables $I_{h/e}^{\pm}$, we perform the analysis on $I'$defined
as

\begin{equation}
I'(B)=\frac{\Phi_{0}}{2\pi A\sin\theta}\frac{\partial I}{\partial B}\label{eq:Iprime}\end{equation}
The normalization is such that the oscillations of $I'(B)$ have the
same amplitude as those in $I(B)$. The reason we use $I'$ is that
the numerical integration can introduce unwanted correlations in the
values of the variables $I_{h/e}^{\pm}$.

\subsection{Drift removal}

We measured cantilevers similar to the ones shown in Fig. \ref{fig:Picture-of-cantilevers}.
We fabricated long $(l\sim400\:\mathrm{\mu m})$ and short $(l\sim100\:\mathrm{\mu m})$
cantilevers, with three different widths ($w=20$, 40 and 60 $\mathrm{\mu m}$).
The signal $\delta f$ was larger for the shortest cantilevers. However,
their noise peformance was very poor, presenting frequency noise with
a power dependence $\sim1/f^{n}$ with $n\approx2$, preventing our
measurements from achieving the thermal noise limit. However, for
the long cantilevers the noise performance of the frequency measurement
did reach the thermal noise limit (see Sect. \ref{sub:PhaseNoise}).

\begin{center}
\begin{figure}[H]
\begin{centering}
\includegraphics[scale=0.6]{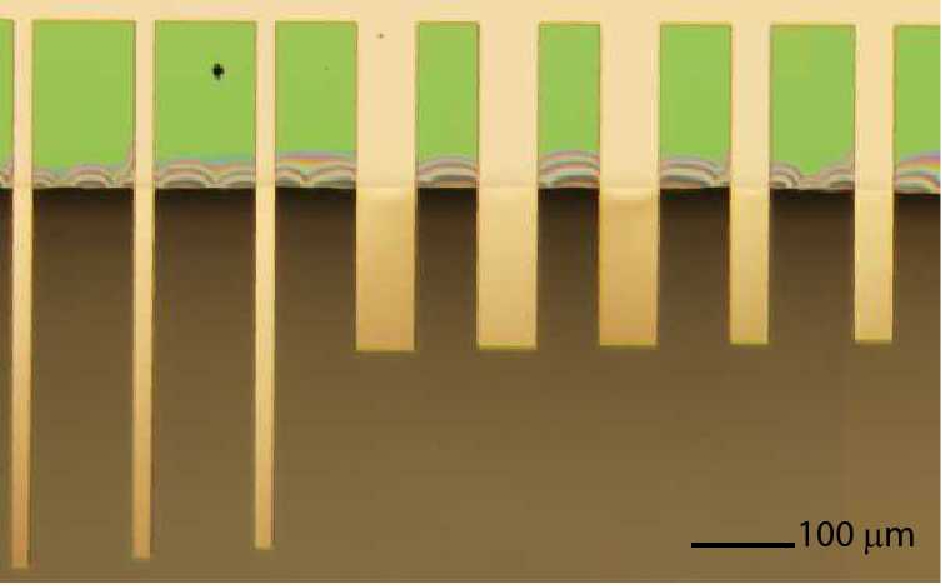}
\par\end{centering}

\caption{\label{fig:Picture-of-cantilevers}Optical micrograph of cantilevers
similar to those used for the measurement of persistent current in
the main text. The ring sample on each cantilever is not visible.}

\end{figure}

\par\end{center}

Raw data of the frequency measurements of two kind of cantilevers
are shown in figures \ref{fig:drift1} and \ref{fig:drift2}. In both
figures we can see a drift in the cantilever's resonance frequency
with time, but with very different characterisitics. In Fig. \ref{fig:drift1}
the Aharonov-Bohm (AB) oscillations can be easily distinguished on
top of a mostly magnetic-field dependent drift. This drift is removed
using MATLAB's local regression algorithm for the LOWESS (Locally
Weighted Scatterplot Smoothing) routine with a first degree polynomial.
We have observed that if we choose the window of the LOWESS routine
to be the equivalent of 5 AB oscillations, the peak of our signal
in its Fourier transform is unchanged by the drift removal process.

For a thermally limited frequency measurement, shorter cantilevers
should have a greater signal-to-noise ratio, but we found that, in
practice, their frequency noise was significantly worse than the longer
cantilevers. Thus, for shorter cantilevers, the frequency measurements
were not thermally limited. This is obvious in figures \ref{fig:drift2}a
and b where it is not possible to discern the AB oscillations at first
sight. In Fig. \ref{fig:drift2}c and d we show the averaged data
and their spectral densities with and without LOWESS drift removal
in order to show that our procedure does not convert a noise shoulder
into a peak.

\begin{center}
\begin{figure}[H]
\centering{}\includegraphics{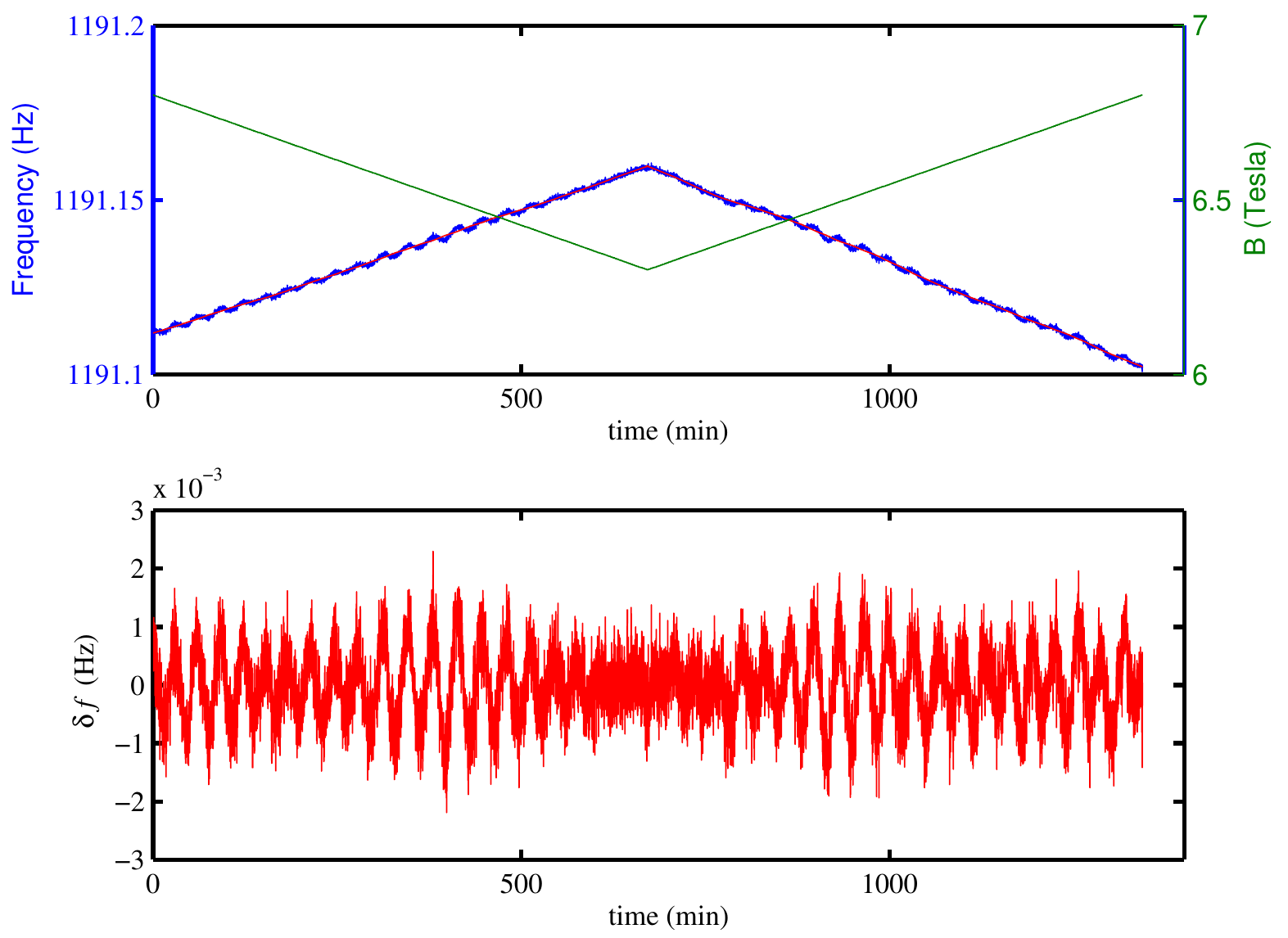}\caption{\label{fig:drift1}(a) Cantilever frequency (blue left axis) as a
function of time as the magnetic field (green right axis) is varied
for sample 6. On top of a magnetic-field-dependent background, the
oscillations due to the persistent current are obvious and the drift
in the cantilever frequency is easily removed by using local regression.
After substracting the drift calculated using LOWESS (red trace) we
obtain the oscillations shown in (b). }

\end{figure}

\par\end{center}

\begin{center}
\begin{figure}[H]
\centering{}\includegraphics{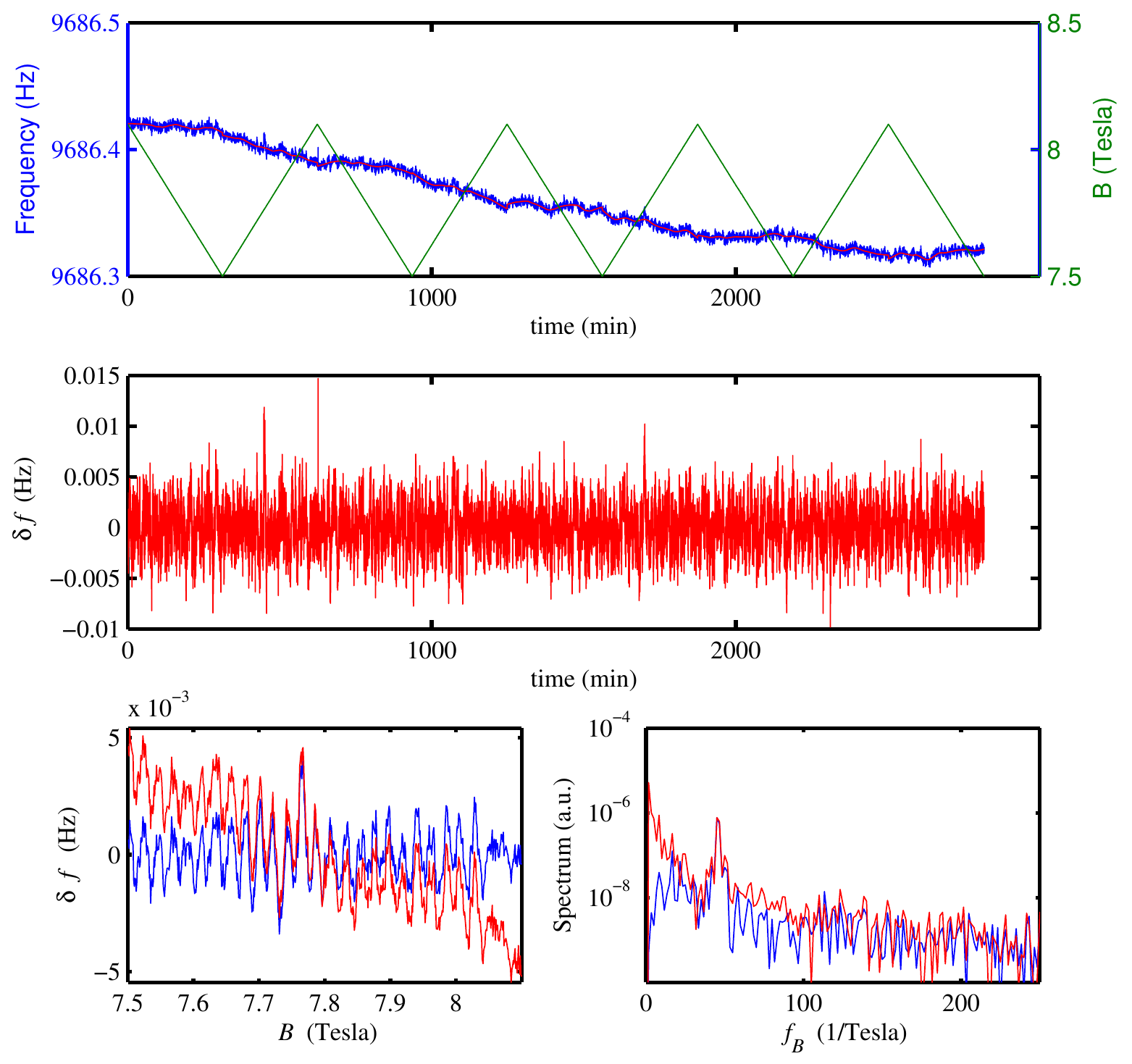}\caption{\label{fig:drift2}(a) Cantilever frequency (blue left axis) as the
magnetic field (green right axis) is varied for sample 2. In this
case, the time dependent background is so large, it is hard to distinguish
the AB oscillations even after removing the drift as shown in (b).
(c) The AB oscillations can be easily distinguished after averaging
all the data with the drift (red) and with the drift substracted (blue).
This proves that the observed oscillations are not an artifact of
the procedure used to remove the drift. (d) Spectral density of (c)
showing that the procedure used to remove the drift does not affect
the amplitude of the observed oscillations.}

\end{figure}

\par\end{center}

\subsection{Hilbert Transform}

The Hilbert transform of a function $u(t)$ is defined as:

\[
\textrm{p.v.}\,\frac{1}{\pi t}\otimes u(t)=H[u(t)]\equiv\hat{u}(t)\]
where p.v. is the Cauchy principal value and $\otimes$ indicates
convolution. The Bedrosian theorem states that the Hilbert transform
acting on the product of two functions $x(t)=A(t)\times h(t)$ can
be written as

\begin{equation}
H[A(t)h(t)]=A(t)H[h(t)]\label{eq:bedrosian}\end{equation}
if the Fourier spectra for $A(t)$ and $h(t)$ are disjoint in frequency
space and if the spectrum for $h(t)$ is concentrated at higher frequencies
than those of $A(t)$. For example, if $h(t)$ is a periodic function
$h(t)=\cos\phi(t)$, $\hat{h}(t)=\sin\phi(t)$. In this regime, then
the following relations hold \citet{Huang}:

\begin{equation}
x(t)=A(t)\cos\phi(t)\label{eq:envelope}\end{equation}

\begin{equation}
A(t)=\textrm{abs}(x(t)+i\hat{x}(t))\label{eq:amplitude}\end{equation}

\begin{equation}
\phi(t)=\textrm{arg}(x(t)+i\hat{x}(t))\label{eq:phase}\end{equation}

For the work described in this paper, the Hilbert transform is used
to analyze two separate data sets. In the first case, the Hilbert
transform is applied to raw interferometer data (i.e. cantilever position
vs. time). This technique allows us to generate densely sampled frequency
vs. time traces, which are used to diagnose the sources of noise in
the measurement. We discuss this technique in section \ref{sub:PhaseNoise}.
In the second case, the Hilbert transform is used to extract the persistent
current quadrature amplitudes $I_{h/e}^{\pm}$ from the persistent
current data. This is discussed in section \ref{sub:amplitude}.

\subsubsection{Phase Noise Analysis\label{sub:PhaseNoise}}

As explained in Ref.\citet{Ania_Science}, the persistent current
measurement is at its core a frequency ($f$) measurement of a driven
cantilever. In order to better understand the specific noise sources
of the measurement system, it is beneficial to have access to the
noise spectral density $S_{f}(\omega)$ of raw frequency data $f(t)$\textcolor{red}{{}
}\[
S_{f}(\omega)=2\lim_{T\rightarrow\infty}\frac{\left|F_{T}(\omega)\right|^{2}}{T}\]
where $F_{T}(\omega)$ is the windowed fourier transform of $f(t)$
\[
F_{T}(\omega)=\intop_{-T/2}^{T/2}dt'f(t)e^{i\omega t'}\]
The factor of 2 comes from the fact that we are only considering single-sided
spectral densities (thus we only consider $\omega>0$ values). For
the work described in the main text, we utitlize a technique involving
the Hilbert transform to measure the frequency $f(t)$. The Hilbert
transform is applied to cantilever position data obtained by an optical-fiber
interferometer.

If the interferometer data contains more than one frequency component,
we can define an {}``instantaneous frequency'' $f(t)=f_{0}+f_{noise}(t)$.
Here $f_{noise}(t)$ is a stochastic variable with zero mean and $f_{0}$
is a constant. We will consider the limit in which the Fourier transform
of the cantilever position is sharply peaked at $f_{0}$ and also
$f_{0}\gg\Delta f$, where $f_{0}$ is the center frequency and $\Delta f$
is the width of the peak. In this limit, the Bedrosian theorem (Eq.
\ref{eq:bedrosian}) holds, and we can use the Hilbert transform to
calculate the cantilever phase vs. time with Eq. \ref{eq:phase}.

\begin{figure}
\begin{centering}
\includegraphics{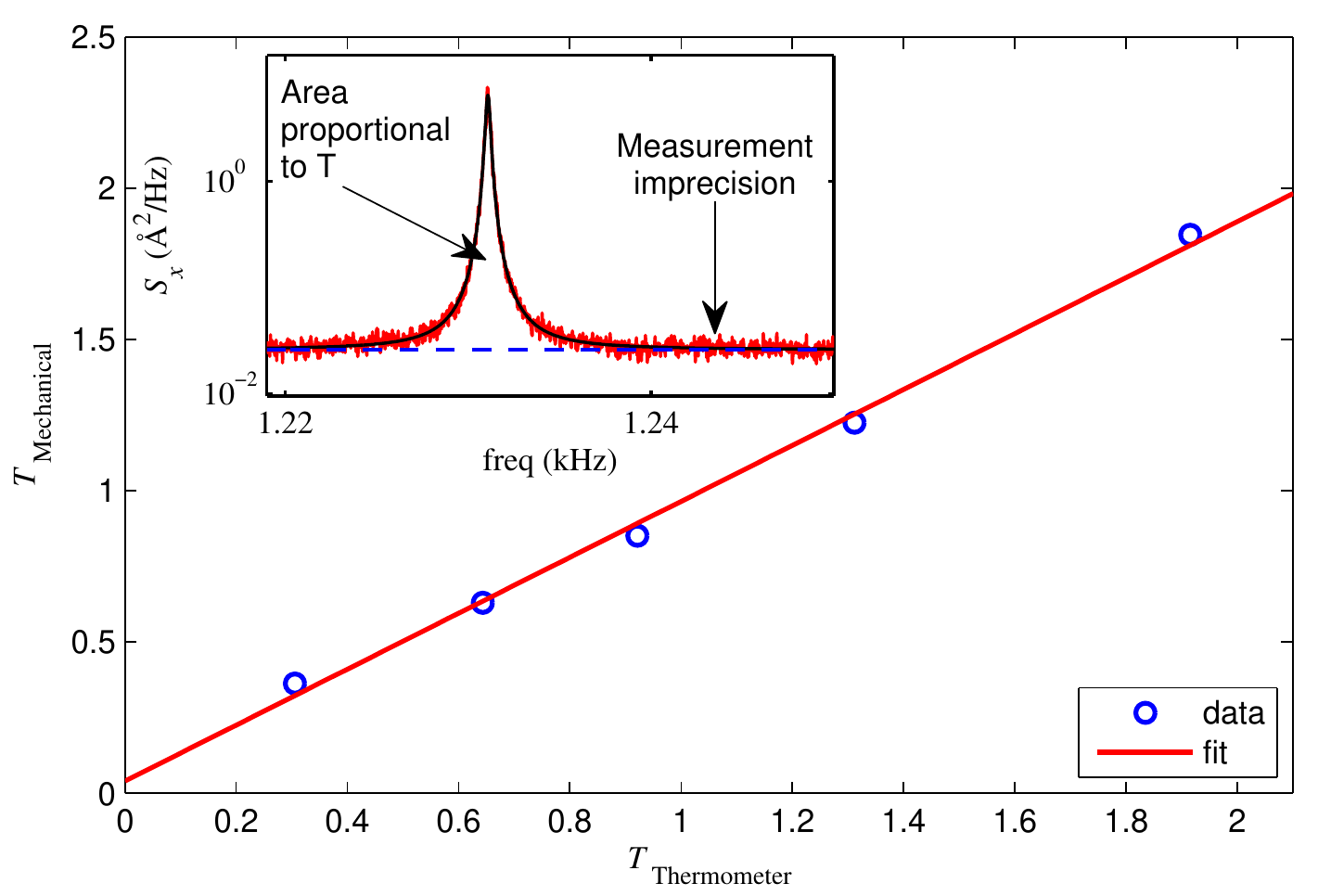}
\par\end{centering}

\centering{}\caption{\label{fig:Thermal-motion}Thermal motion of the cantilever in units
of temperature (Eq. \ref{eq:Equipartitino}) as a function of the
fridge temperature. This motion is measured as a voltage at the output
of the photodetector and converted into displacement units (inset).
The spectral density of the voltage has two constributions: a white
background that consists of the measurement imprecision $S_{x}^{\mathrm{imp}}$
coming from the voltage noise of our photodetector (dashed blue line
in the inset) , and a lorentzian curve that consists of the thermal
motion of the cantilever. Measuring the detector's voltage fluctuations
at different temperatures allows us to calibrate the intereferometer
response. }

\end{figure}

We convert the phase versus time $\phi(t)$ of the cantilever's motion
to an instantaneous frequency $f(t)$ via the relationship:

\begin{equation}
2\pi f(t)=\frac{d\phi}{dt}\label{eq:instantf}\end{equation}
The frequency noise spectrum is related to the phase noise spectrum
by $S_{f}=(\omega/2\pi)^{2}S_{\phi}$ by Eq. (\ref{eq:instantf}).
We can then compare the measured $S_{f}$ to theoretical predictions
for the frequency noise of a driven limit-cycle oscillator subject
to a white force noise and a white displacement noise in the detection
\citet{Albrecht}

\begin{eqnarray}
S_{f,\,\mathrm{thermal}}(\omega) & = & \frac{f_{0}k_{B}T}{\pi kQx_{\mathrm{tip}}^{2}}\nonumber \\
S_{f,\,\mathrm{detector}}(\omega) & = & \frac{2S_{x}^{\mathrm{imp}}}{x_{\mathrm{tip}}^{2}}\left(\frac{\omega}{2\pi}\right)^{2}\label{eq:SffAlbrecth}\end{eqnarray}
where $k$ is the resonator spring constant, $Q$ is the resonator
quality factor, $x_{\mathrm{tip}}$ is the resonator displacement
amplitude, and $S_{x}^{\mathrm{imp}}$ is the displacement noise of
the detector. The imprecision $S_{x}^{\mathrm{imp}}$ in the measurement
of the cantilever's motion arises from fluctuations both in the laser
used to monitor the cantilever and in the detector used to measure
the laser signal. The dominant noise source in our case is the electronic
noise from the photoreciver. In order to convert the voltage at the
output of the photodetector into cantilever motion, we measure this
voltage when the cantilever is only excited by a thermal force. The
magnitude of the cantilever's thermal motion can be computed from
the equipartition theorem which states that at thermal equilibrium
\begin{equation}
k\left\langle x^{2}\right\rangle =k_{B}T\label{eq:Equipartitino}\end{equation}
The spectrum of the voltage $S_{V}$ at the output of the photodectector
has the shape of lorentzian curve on top of an offset (inset of Figure
\ref{fig:Thermal-motion}). This offset consists of the measurement
imprecision $S_{x}^{\mathrm{imp}}$. By measuring $S_{V}$ at different
temperatures (Figure \ref{fig:Thermal-motion}), we can properly convert
the volts at the output of the detector into displacement of the cantilever
and calculate the imprecision $S_{x}^{\mathrm{imp}}$ of our detector.

\begin{figure}
\begin{centering}
\includegraphics{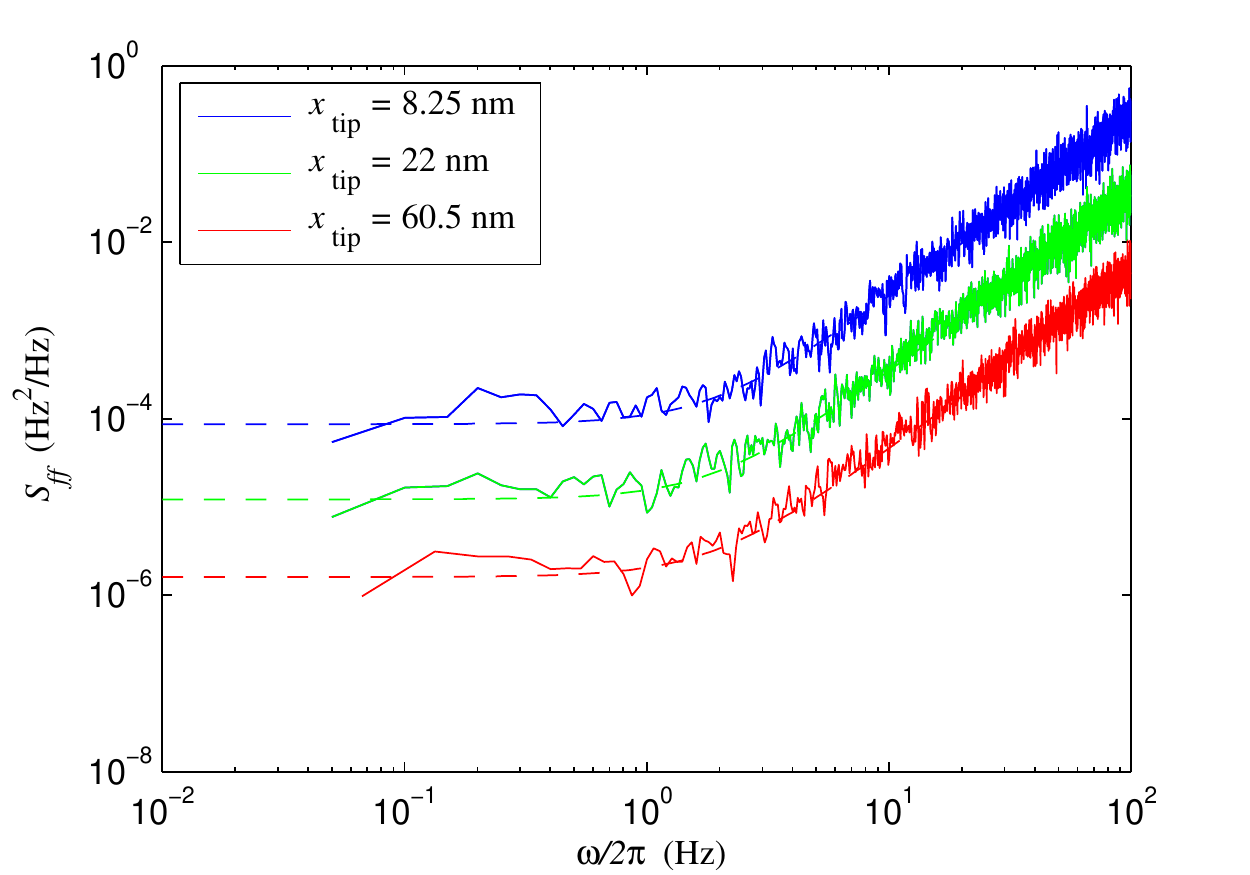}
\par\end{centering}

\centering{}\caption{\label{fig:phasenoise32}Frequency noise for sample 4. The solid lines
represent the frequency noise spectrum data obtained using the Hilbert
transform technique described above at various oscillator drive strengths.
The dashed lines are the theoretical noise curves calculated with
Eq. \ref{eq:SffAlbrecth} (no fit parameters). There are two major
contributions to the noise: thermal fluctuations driving the cantilever,
and the intrinsic detector noise. The detector noise was extracted
from the actual interferometer noise. For this cantilever, $Q=5700$,
$T=320\:\mathrm{mK}$, $f_{0}=966\:\mathrm{Hz}$ and $k_{L}=4.0\times10^{-5}\:\mathrm{N/m}.$}

\end{figure}

\begin{figure}
\begin{centering}
\includegraphics{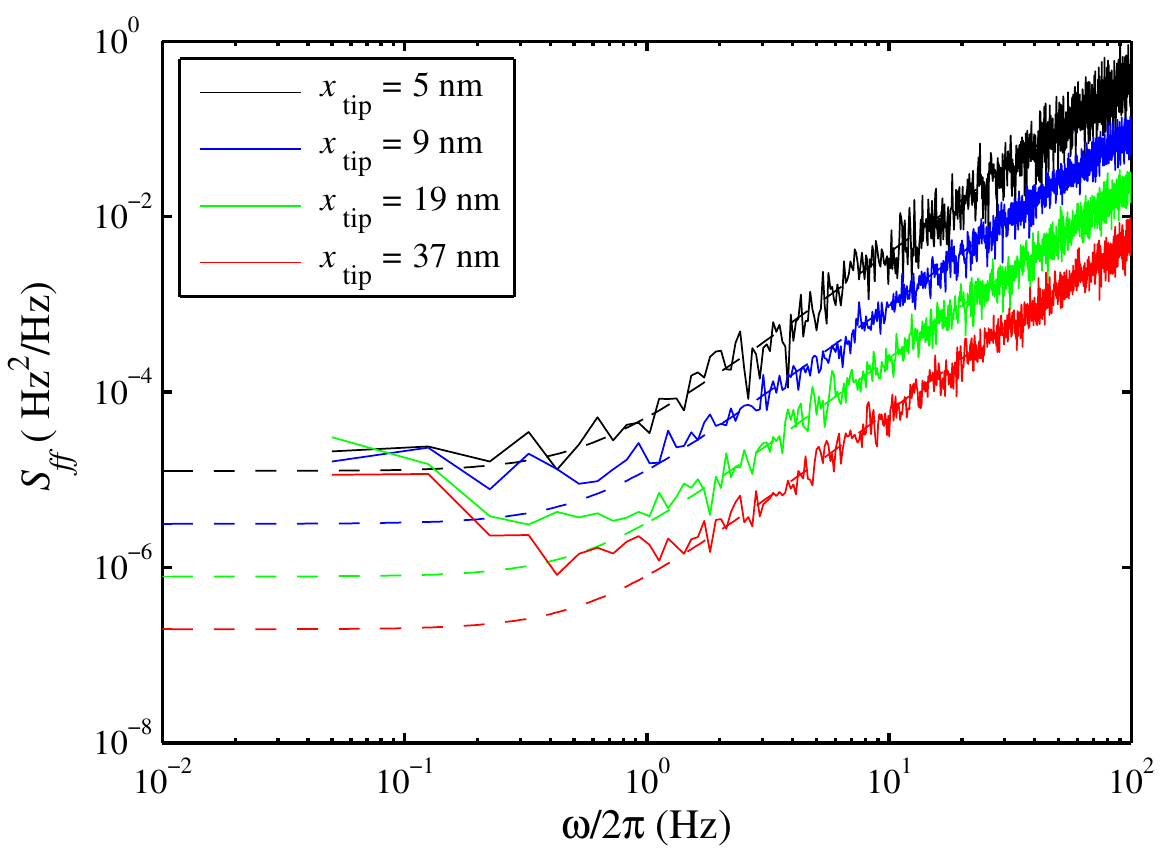}
\par\end{centering}

\centering{}\caption{\label{fig:phasenoise04}Frequency noise for sample 2. The solid lines
represent the frequency noise spectrum data obtained using the Hilbert
transform technique described above at various oscillator drive strengths.
The dashed lines are the theoretical noise curves calculated with
Eq. \ref{eq:SffAlbrecth} (no fit parameters). This sample exhibited
strongly nonlinear behavior and its low-frequency noise deviates from
the theoretical curves. For this cantilever, $Q=35000$, $T=320\:\mathrm{mK}$,
$f_{0}=11827\:\mathrm{Hz}$ and $k_{L}=1.7\times10^{-3}\:\mathrm{N/m}.$}

\end{figure}

In these measurements, for some cantilevers, we notice a strong deviation
from the prediction at low frequencies. This low-frequency behavior
seems to correlate with the amount of mechanical nonlinearity in the
cantilever (determined, e.g., by a non-Lorentzian resonance for large
$x_{\mathrm{tip}}$). For samples 3-8, there was very little nonlinearity
and the frequency noise behaves as expected (Fig. \ref{fig:phasenoise32}).
For samples 1 and 2, we noticed a large amount of mechanical nonlinearity
and also a large amount of low-frequency noise (Fig. \ref{fig:phasenoise04}).

\subsubsection{Calculating the Persistent Current Amplitude\label{sub:amplitude}}

The goal of the work in the main text is to measure the statistics
of the quadrature amplitudes of the AB oscillations $I_{h/e}^{\pm}$.
This amplitude slowly changes due to the magnetic field that penetrates
the metal of the ring \citet{Ginossar2010}. Since our data is dominated
by the first harmonic of Eq. \ref{eq:Ifourierseries} (no higher harmonics
are visible in the data) the normalized derivative of the current
$I'$ (Eq. \ref{eq:Iprime}) takes the form:

\begin{eqnarray}
I'(B) & = & I_{h/e}^{+}\cos(2\pi p\frac{\phi}{\Phi_{0}})-I_{h/e}^{-}\sin(2\pi p\frac{\phi}{\Phi_{0}})\label{eq:pcdata}\\
 & = & I_{h/e}(B)\cos\left(2\pi p\frac{BA\sin\theta}{\Phi_{0}}+\alpha(B)\right)\end{eqnarray}
where $I_{h/e}(B)$ is the amplitude of the current and $\cos\left(2\pi p\frac{BA\sin\theta}{\Phi_{0}}-\alpha(B)\right)$
represents the AB oscillations.

\begin{figure}
\centering{}\includegraphics{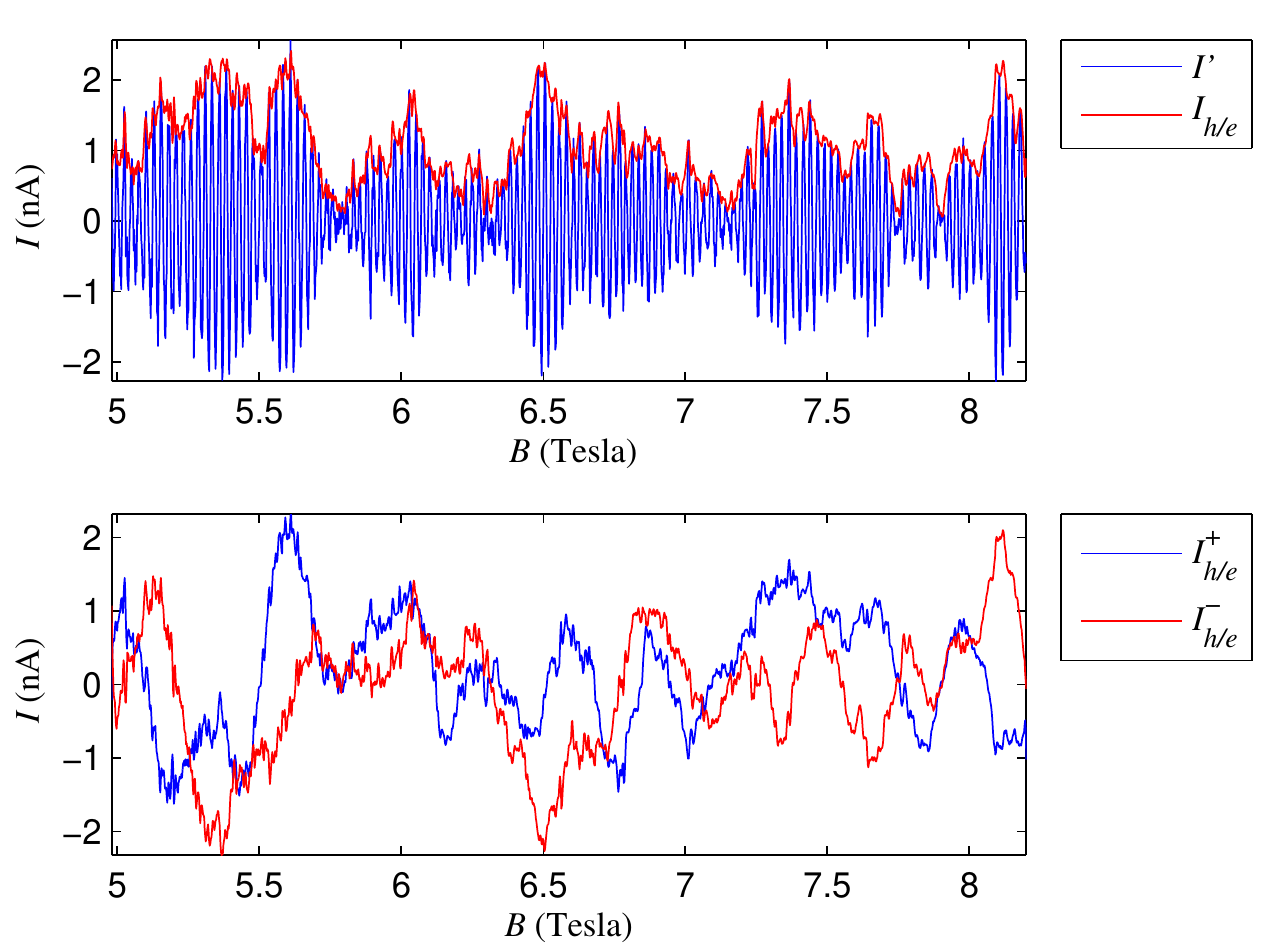}\caption{\label{fig:HilbTrans} (a) Hilbert transform for the data from sample
6. From this trace we obtain $I_{h/e}(B)$ and $\alpha$ (not shown).
(b) Using the data from (a) we can infer the values for $I_{h/e}^{+}$
and $I_{h/e}^{-}$.}

\end{figure}

The correlation field $B_{c}$ sets the field scale over which $I_{h/e}(B)$
changes. In the limit that $B_{c}$ is {}``large enough'' ($B_{c}\gg\Phi_{0}/A$),
$I_{h/e}(B)$ is a slowly-varying function compared to $\cos(2\pi p\frac{BA\sin\theta}{\Phi_{0}}-\alpha)$
and the Bedrosian theorem (\ref{eq:bedrosian}) holds. The AB frequency
is also sharply peaked in Fourier space, with a well-defined frequency
given by the dimensions of the ring. Thus, comparing (\ref{eq:amplitude})
and (\ref{eq:pcdata}), we can obtain $I_{h/e}(B)$ and $\alpha$
using the Hilbert transform. This also allows us to determine the
quadrature variables $I_{h/e}^{+}=I_{h/e}\cos(\alpha)$ and $I_{h/e}^{-}=I_{h/e}\sin(\alpha)$.
This technique is illustrated in Fig. \ref{fig:HilbTrans}.

\subsection{Definition of the cumulants and statistical moments}

For the purpose of this manuscript, it is convenient to have a consistent
definition of the various moments and cumulants. The raw moments $\mu'_{r}$
and central moments $\mu{}_{r}$ of a stochastic variable $x$ are
defined as \begin{eqnarray}
\mu'_{r} & = & \left\langle x^{r}\right\rangle \label{eq:moments}\\
\mu{}_{r} & = & \left\langle \left(x-\mu'_{1}\right)^{r}\right\rangle \nonumber \end{eqnarray}
The cumulants $\left\langle \left\langle I^{n}\right\rangle \right\rangle $
are most easily defined in terms of the central moments:\begin{eqnarray}
\left\langle \left\langle I^{2}\right\rangle \right\rangle  & = & \mu_{2}\nonumber \\
\left\langle \left\langle I^{3}\right\rangle \right\rangle  & = & \mu_{3}\nonumber \\
\left\langle \left\langle I^{4}\right\rangle \right\rangle  & = & \mu_{4}-3\mu_{2}^{2}\label{eq:cumulantdefinition}\\
\left\langle \left\langle I^{5}\right\rangle \right\rangle  & = & \mu_{5}-10\mu_{3}\mu_{2}\nonumber \\
\left\langle \left\langle I^{6}\right\rangle \right\rangle  & = & \mu_{2}-15\mu_{4}\mu_{2}-10\mu_{3}^{2}+30\mu_{2}^{3}\nonumber \end{eqnarray}

\subsection{Finite sample statistics}

In Ref.\citet{Ginossar2010} it was mentioned that in the presence
of an additional large in-plane magnetic field $B_{m}$ penetrating
the metal ring, the effective disorder of the ring changes, implying
that averaging over magnetic fi{}eld is equivalent to an ensemble
average\citet{Lee1985a}. However, our finite magnetic field range
means that in practice, we have a finite number of realizations from
this ensemble. In order to estimate the statistical uncertainty in
our estimates of the cumulants due to this finite sample size, we
use the results of Ref.\citet{ErgodicityProblem}.

\begin{figure}
\begin{centering}
\includegraphics{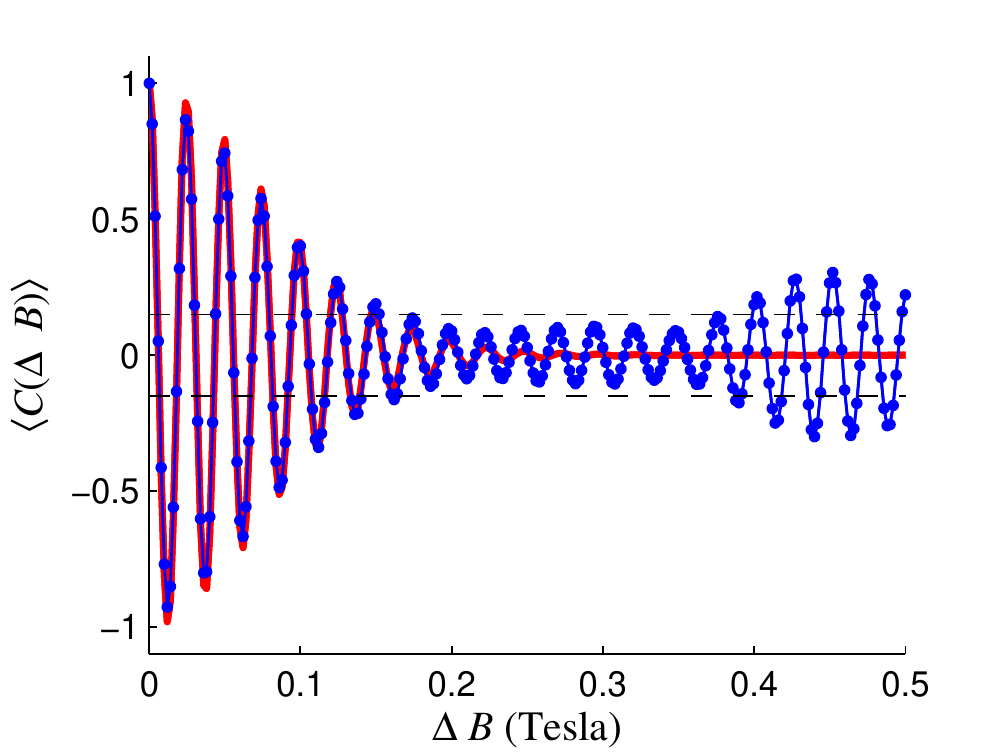}
\par\end{centering}

\centering{}\caption{\label{fig:Persistent-current-autocorrelation}Persistent current
autocorrelation for sample 6. Blue represents the autocorrelation
of the data shown in Fig. \ref{fig:CL62} and red is the fit using
Ref.\citet{Ginossar2010}. The estimated correlation field in this
case is $B_{c}=32\:\mathrm{mT}.$ In order to estimate the standard
error in the autocorrelation (dashed horizontal lines), we assume
that this is approximatly given by the standard error for the correlation
coefficient $1/\sqrt{N_{\text{\textrm{\textrm{eff}}}}}\approx0.15$.
As it can be seen in the figure, fluctuations of this order are present
at large $\Delta B\gg B_{c}$, where the autocorrelation of an infinitely
large data set would be expected to vanish.}

\end{figure}

The statistical uncertainty of the cumulants ($\left\langle \left\langle I^{n}\right\rangle \right\rangle $)
due to a finite sample size can be expressed in terms of the normalized
correlation function $C(x)$ defined as\begin{eqnarray*}
C\left(\frac{B-B'}{B_{c}}\right) & = & \frac{\left\langle I(B)I(B')\right\rangle }{\left\langle I^{2}\right\rangle }\end{eqnarray*}
where $C$ decays from $C(0)=1$ to $C(\infty)=0$, and $B_{c}$ is
the correlation field, which sets a rough order of magnitude over
which the persistent current is correlated. Expressions for $C$ are
provided in Ref.\citet{Ginossar2010}. The typical value of a cumulant
calculated from a finite ensemble is given by \begin{eqnarray*}
\left\langle \left\langle \left\langle I^{n}\right\rangle \right\rangle ^{2}\right\rangle _{B} & = & \left\langle \left\langle I^{n}\right\rangle \right\rangle ^{2}+a_{n}\frac{B_{c}}{B_{\text{span}}}\left\langle \left\langle I^{2}\right\rangle \right\rangle ^{n}\end{eqnarray*}
where $a_{n}=n!\int_{-\infty}^{\infty}[C(x)]^{n}dx$, and $B_{\text{span}}$
is the total magnetic field range over which data is taken. The theoretical
prediction for the higher order cumulants of the persistent current
quadratures is \[
\left\langle \left\langle I^{n}\right\rangle \right\rangle \sim\frac{I_{\mathrm{typ}}^{n}}{g^{n-2}}\]
for $n>2$, where the typical current is defined as $I_{\mathrm{typ}}=\sqrt{\left\langle I^{2}\right\rangle }$\citet{Houzet2010}.
In our case $g\sim10^{4}$ and thus we expect a gaussian distribution
for $I_{h/e}^{\pm}$ and the statistical error is thus given by \begin{equation}
\delta\left\langle \left\langle I^{n}\right\rangle \right\rangle =\sqrt{a_{n}\frac{B_{c}}{B_{\text{span}}}\left\langle I^{2}\right\rangle ^{n}}\label{eq:StatError}\end{equation}
For an estimate of this term, we can consider the case for a spinless
electron at $T=0$ and large $B_{m}$ (so that we can ignore the Cooperon
contribution) and only include the first harmonic of the PC. In this
case \begin{equation}
C(x)=\left(1+|x|+\frac{x^{2}}{3}\right)e^{-|x|}\label{eq:K(x)}\end{equation}
For this simplified case, the first 4 coefficients are $a_{2}=7$,
$a_{3}=16.61$, $a_{4}=56.48$, and $a_{5}=249.55$. Experimental
values of the cumulants of the persistent currents are considered
sound only if they comfortably exceed the systematic error $\sqrt{a_{n}\frac{B_{c}}{B_{\text{span}}}\left\langle I^{2}\right\rangle ^{n}}$
(Eq. \ref{eq:StatError}). The actual correlation function for our
case is more complicated than Eq. \ref{eq:K(x)} and is a function
of $T$ and $L_{SO}$. It can be numerically implemented using Ref.\citet{Ginossar2010}.
A fit to the correlation function $C(B/B_{c},T,L_{SO})$ of our data
is shown in Fig. \ref{fig:Persistent-current-autocorrelation}.%
\footnote{The correlation function is a also a function of the Zeeman energy,
but at the large magnetic fields used in this experiment, this dependence
is negligible.%
}

We can also define an effective number of samples based on the expected
statistical uncertainty for the cumulants of an ensemble of $N_{\text{eff}}$
samples \citet{kendall}:\[
\delta\left\langle \left\langle I^{n}\right\rangle \right\rangle =\sqrt{\frac{n!\left\langle \left\langle I^{2}\right\rangle \right\rangle ^{n}}{N_{\text{eff}}}}.\]
By comparing the latter with Eq. \ref{eq:StatError}, we can then
define $N_{\text{eff}}$ as\begin{equation}
N_{\text{\text{n,eff}}}=\frac{n!B_{\text{span}}}{a_{n}B_{c}}.\label{eq:effectiven}\end{equation}

\subsection{Finite signal to noise}

Another source of error in our estimate of the cumulants is the finite
sensitivity of our setup. In order to estimate this error, we have
to calcualte the appropriate expression for error propagation in our
analysis. Our case is somewhat confusing, since $x_{i}$ itself is
an stochastic variable, so really we have $x_{i,meas}=x_{i}+\delta x_{i}$.
First let's consider the general equation for the error propagation
for a general function $f(x_{1},x_{2}...,x_{N})$. If $x_{i}$ is
known only to within an error $\delta x_{i}$ then the error in $f$
is given by \begin{equation}
\delta^{2}f=\sum_{i}\left(\frac{\partial f}{\partial x_{i}}\delta x_{i}\right)^{2}\label{eq:properrorgeneral}\end{equation}
 Let's first consider the case of the central moment as cumulants
are easily definined in terms of them:\[
f(x_{i})=\mu_{r}=\frac{1}{N}\sum_{i}\left(x_{i}-\mu'_{1}\right)^{r}\]
Then \begin{eqnarray}
\left\langle \delta^{2}\mu_{r}\right\rangle  & = & \left(\frac{r}{N}\right)^{2}\sum_{i}\left(x_{i}-\mu'_{1}\right)^{2r-2}\left\langle \delta^{2}x_{i}\right\rangle \nonumber \\
 & = & \frac{r^{2}}{N}\mu_{2r-2}\left\langle \delta^{2}x\right\rangle \label{eq:properror1}\end{eqnarray}
There are two assumptions used for Eq. \ref{eq:properror1}. First,
we assume that the noise $\delta x_{i}$ from the different magnetic
field points (labeled with the index $i$) are uncorrelated. This
\textbf{does not }mean that the different $x_{i}$'s are uncorrelated.
The second assumption is that $\delta^{2}x_{i}$ can be replaced by
an average $\delta^{2}x$. Although not completely accurate, since
the sensitivity of our measurement is lower at lower magnetic fields,
we did compensate for the loss of sensitivity by averaging longer
at low magnetic fields.

The expressions for the second and third cumulants are the same as
for the second and third central moments. Thus, Eq. \ref{eq:properror1}
gives their measurement uncertainty. For the forth, fifth and sixth
cumulant we can use the following heuristic approach\citet{kendall}.
We start with the definition of the cumulants (Eq. \ref{eq:cumulantdefinition}),
from which the following expressions are derived: \begin{eqnarray*}
\delta\left\langle \left\langle I^{4}\right\rangle \right\rangle  & = & \delta\mu_{4}-6\mu_{2}\delta\mu_{2}\\
\delta\left\langle \left\langle I^{5}\right\rangle \right\rangle  & = & \delta\mu_{5}-10\mu_{2}\delta\mu_{3}-10\mu_{3}\delta\mu_{2}\\
\delta\left\langle \left\langle I^{6}\right\rangle \right\rangle  & = & \delta\mu_{6}-15\mu_{4}\delta\mu_{2}-15\mu_{2}\delta\mu_{4}-20\mu_{3}\delta\mu_{3}+90\mu_{2}^{2}\delta\mu_{2}\end{eqnarray*}
Then, for example, for the fourth cumulant: \begin{equation}
\left\langle \delta^{2}\left\langle \left\langle I^{4}\right\rangle \right\rangle \right\rangle =\delta^{2}\mu_{4}+36\mu_{2}^{2}\delta^{2}\mu_{2}-12\mu_{2}\left\langle \delta\mu_{4}\delta\mu_{2}\right\rangle \label{eq:properror2}\end{equation}
The covariance $\left\langle \delta\mu_{4}\delta\mu_{2}\right\rangle $
is not necessarily zero and it can be derived using a similar approach
as the one used to derive Eq. \ref{eq:properror1}:\[
\left\langle \delta\mu_{r}\delta\mu_{s}\right\rangle =\frac{rs}{N}\mu_{r+s-2}\left\langle \delta^{2}x\right\rangle \]
 Similar expressions to Eq. \ref{eq:properror2} for $\left\langle \left\langle I^{5}\right\rangle \right\rangle $
and $\left\langle \left\langle I^{6}\right\rangle \right\rangle $
can be derived.

\subsection{Data}

Following, we show a table with all the estimated cumulants and estimated
errors for the cumuluants of all 8 samples. The cumulants of both
$I^{+}$ and $I^{-}$ have been combined for each of the rings. In
order to account for variations between rings the different cumulants
are normalized by the variance $\left\langle \left\langle I^{2}\right\rangle \right\rangle $;
thus we define a normalized cumulant $\kappa_{n}\equiv\left\langle \left\langle I^{n}\right\rangle \right\rangle /\left\langle \left\langle I^{2}\right\rangle \right\rangle ^{n/2}$.

\begin{table}[H]
\begin{centering}
\begin{tabular}{|c|c|c|c|c|c|}
\hline
Sample \# & $r$ (nm) & $B_{c}$ (mT) & $N_{\mathrm{2,eff}}$ & $\left\langle \left\langle \left(I^{\pm}\right)^{2}\right\rangle \right\rangle (\mathrm{nA^{2}})$ & $\kappa_{3}$\tabularnewline
\hline
\hline
1 & 296 & $16\pm1$ & 63 & $0.46\times(1\pm0.18\pm0.02)$ & $0.065\pm0.28\pm0.05$\tabularnewline
\hline
2 & 296 & $25\pm1.5$ & 43 & $0.37\times(1\pm0.22\pm0.02)$ & $0.23\pm0.34\pm0.04$\tabularnewline
\hline
3 & 448 & $9\pm1$ & 55 & $0.0031\times(1\pm0.2\pm0.03)$ & $0.21\pm0.31\pm0.08$\tabularnewline
\hline
4 & 448 & $17\pm1$ & 34 & $0.0063\times(1\pm0.24\pm0.02)$ & $-0.25\pm0.4\pm0.07$\tabularnewline
\hline
5 & 296 & $22\pm1$ & 56 & $0.597\times(1\pm0.19\pm0.01)$ & $0.18\pm0.29\pm0.03$\tabularnewline
\hline
6 & 296 & $32\pm2$ & 44 & $0.67\times(1\pm0.21\pm0.01)$ & $0.001\pm0.34\pm0.02$\tabularnewline
\hline
7 & 296 & $20\pm1$ & 81 & $0.26\times(1\pm0.16\pm0.01)$ & $0.083\pm0.24\pm0.03$\tabularnewline
\hline
8 & 418 & $11\pm1$ & 36 & $0.022\times(1\pm0.23\pm0.02)$ & $-0.05\pm.37\pm0.05$\tabularnewline
\hline
\end{tabular}
\par\end{centering}

\begin{centering}
\begin{tabular}{|c|c|c|c|}
\hline
Sample \# & $\kappa_{4}$ & $\kappa_{5}$ & $\kappa_{6}$\tabularnewline
\hline
\hline
1 & $-0.052\pm0.51\pm0.1$ & $-0.12\pm1.1\pm0.25$ & $-0.91\pm2.6\pm0.47$\tabularnewline
\hline
2 & $-0.0271\pm0.62\pm0.07$ & $-0.83\pm1.3\pm0.2$ & $-0.93\pm3.1\pm0.41$\tabularnewline
\hline
3 & $0.29\pm0.58\pm0.18$ & $-0.05\pm1.23\pm0.5$ & $-0.84\pm2.9\pm1.13$\tabularnewline
\hline
4 & $0.56\pm0.75\pm0.16$ & $0.27\pm1.58\pm0.5$ & $-0.043\pm3.7\pm1.32$\tabularnewline
\hline
5 & $-0.25\pm0.54\pm0.04$ & $-0.17\pm1.15\pm0.14$ & $0.36\pm2.7\pm0.28$\tabularnewline
\hline
6 & $-0.61\pm0.637\pm0.03$ & $-0.05\pm1.35\pm0.09$ & $2.23\pm3.15\pm0.18$\tabularnewline
\hline
7 & $0.065\pm0.45\pm0.07$ & $0.51\pm0.95\pm0.23$ & $2.85\pm2.23\pm0.45$\tabularnewline
\hline
8 & $0.2\pm0.68\pm0.086$ & $0.12\pm1.4\pm0.23$ & $-2.1\pm3.35\pm0.58$\tabularnewline
\hline
\end{tabular}
\par\end{centering}

\caption{\label{tab:cumulant} Higher order moments of the quadrature amplitudes
of the persistent current for all the measured samples. The first
error indicates the standard error due to finite-size sample and the
other represents uncertainty from noise in the frequency measurement.
$N_{2,\text{eff}}$ indicates the effective number of samples for
a given ring using Eq. \ref{eq:effectiven} multiplied by 2 as we
use both quadratures $I^{\pm}$ to calculate the cumulants.}

\end{table}

\section{Measurement diagnostics}

We performed a set of diagnostic measurements similar to those described
in Ref.\citet{Ania_Science}. Specifically, we measured the effects
of the readout laser power on the measured frequency, we compared
the extracted current using the two first modes of the cantilever
motion and we checked the effects of cantilever oscillation amplitude.

Figures \ref{fig:powerdep1} and \ref{fig:powerdep2} show the effect
of varying the laser power upon the persistent current data measured
on two different cantilevers for different incident laser powers.
For different cantilevers, we observed two qualitatively different
types of dependence upon the laser power. We believe this difference
is due to the different widths of the cantilevers. The data of samples
1-4 where taken with an incident power of 10 nW. However, for samples
5-7, the data was taken with an laser incident power of 3 nW as they
presented a stronger power dependence. Sample 8 was taken with a laser
power of 5 nW in a previous cooldown (Ref.\citet{Ania_Science}).

\begin{table}[H]
\centering{}\begin{tabular}{|c|c|c|c|c|c|}
\hline
Sample \# & $r$ (nm) & $w$ (nm) & $t$ (nm) & $l_{\mathrm{cantilever}}$ ($\mathrm{\mu m})$ & $w_{\mathrm{cantilever}}$ ($\mathrm{\mu m})$\tabularnewline
\hline
\hline
1 & 296 & 90 & 115 & 114 & 40\tabularnewline
\hline
2 & 296 & 90 & 115 & 126 & 40\tabularnewline
\hline
3 & 448 & 90 & 115 & 395 & 40\tabularnewline
\hline
4 & 448 & 90 & 115 & 398 & 40\tabularnewline
\hline
5 & 296 & 90 & 115 & 370 & 20\tabularnewline
\hline
6 & 296 & 85 & 115 & 359 & 20\tabularnewline
\hline
7 & 296 & 95 & 115 & 352 & 20\tabularnewline
\hline
8 & 418 & 85 & 90 & 438 & 60\tabularnewline
\hline
\end{tabular}\caption{\label{tab:SampleTable}Sample parameters. All the measurements were
done at a 45-degree angle between the cantilever and ring plane and
the magnetic field. All the cantilevers had a thickness of 118 nm
except for sample 8 which had a thickness of $340\text{ nm}$. Sample
8 was measured in a separate cooldown. }

\end{table}

\begin{figure}[H]
\begin{centering}
\includegraphics[clip]{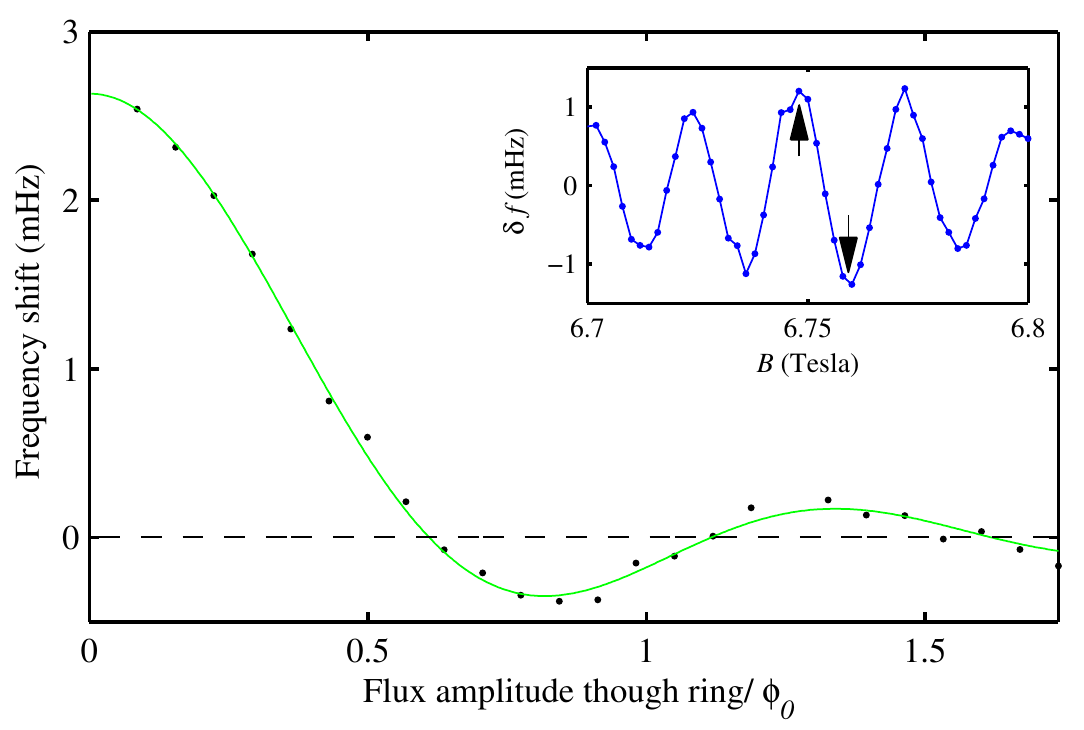}
\par\end{centering}

\centering{}\caption{\label{fig:Cantilever-drive-test}Cantilever drive test on sample
5 where we measure the accuracy of Eq. \ref{eq:deltaf}. The cantilever
amplitude is plotted on the $x$-axis in terms of the amplitude of
the flux modulation $\phi_{ac}/\phi_{0}$ through the ring produced
by the cantilever motion. Data points reperesent the difference in
cantilever frequency shift for the two field values indicated in the
inset. The solid curve is a fit using Eq. \ref{eq:deltaf} (with $p=1$)
and $r=301\pm2\,\mathrm{nm}$, consistent with the measured radius
and linewidth of our ring (see table \ref{tab:SampleTable}). Inset:
the arrows indicate two field values at which measurements of the
cantilever frequency shift were performed as a function of cantilever
amplitude. }

\end{figure}

In Fig. \ref{fig:secondmode} we demonstrate that the inferred current
is the same whether the cantilever's first or second flexural mode
is used, indicating that persistent current is independent of excitation
frequency of the cantilever. Finally, in Fig. \ref{fig:Cantilever-drive-test}
we show that the cantilever's frequency depends upon the amplitude
of its motion as would be expected if the persistent current remains
in its equilibrium state (Eq. \ref{eq:deltaf}). To generate the data
shown in Fig. \ref{fig:Cantilever-drive-test}, the resonant frequency
of cantilever 5 was measured at two different magnetic fields (indicated
in the inset figure with two arrows) as a function of cantilever amplitude
$x_{\mathrm{tip}}$. Then, the two measured $\delta f$ were substracted
in order to remove any kind of amplitude-dependent change in the cantilevers
resonance frequency. The cantilever was excited in its first flexural
mode.

\begin{center}
\begin{figure}[H]
\begin{centering}
\includegraphics{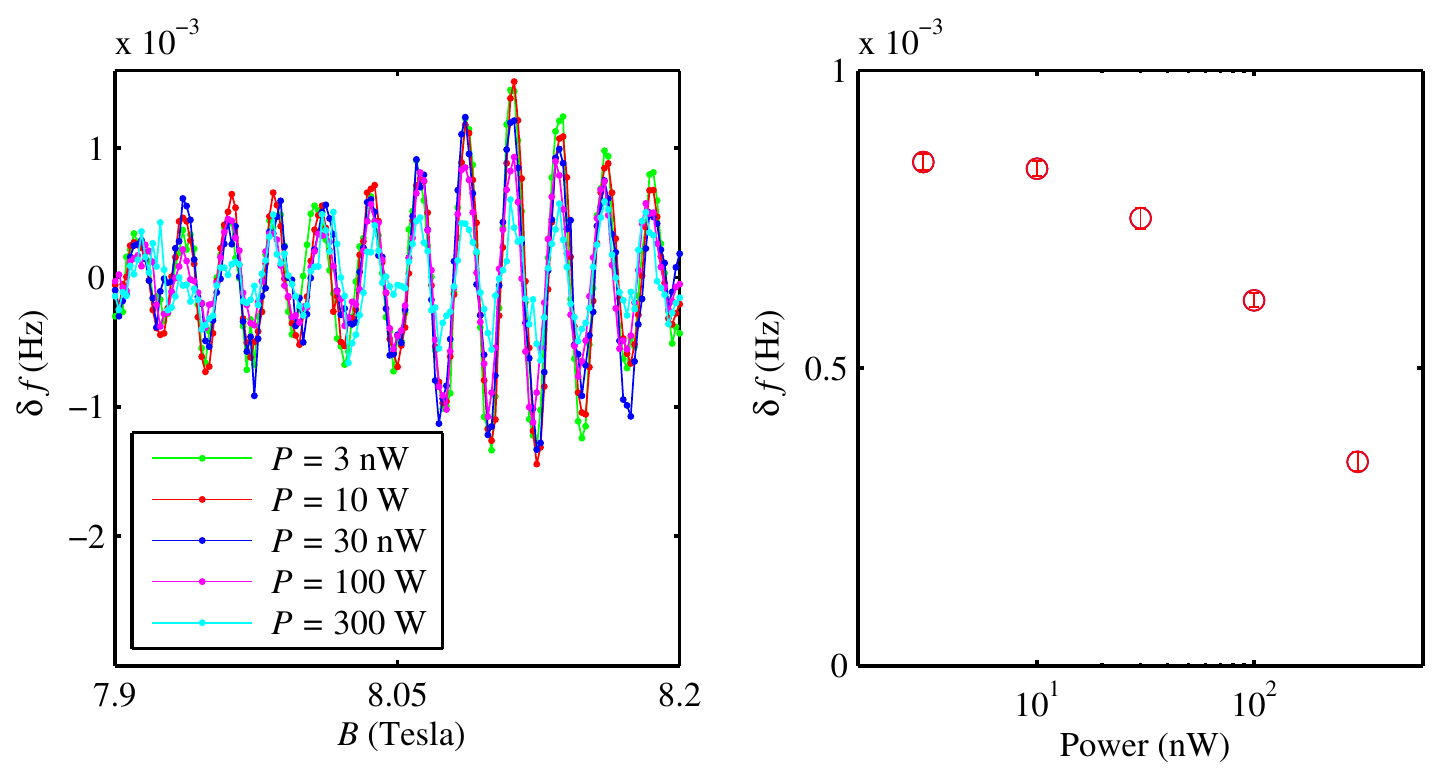}
\par\end{centering}

\centering{}\caption{\label{fig:powerdep1}Left panel: Change in frequency versus magnetic
field for a series of laser powers incident on the cantilever. This
data was taken with sample 6. The drive was the same for all the traces.
Right panel: The mean amplitude of each of the traces from the left
panel, plotted versus laser power. For the data of samples 5, 6 and
7 shown in the main paper, $3\text{ nW}$ of laser power was used.}

\end{figure}

\par\end{center}

\begin{center}
\begin{figure}[H]
\centering{}\includegraphics{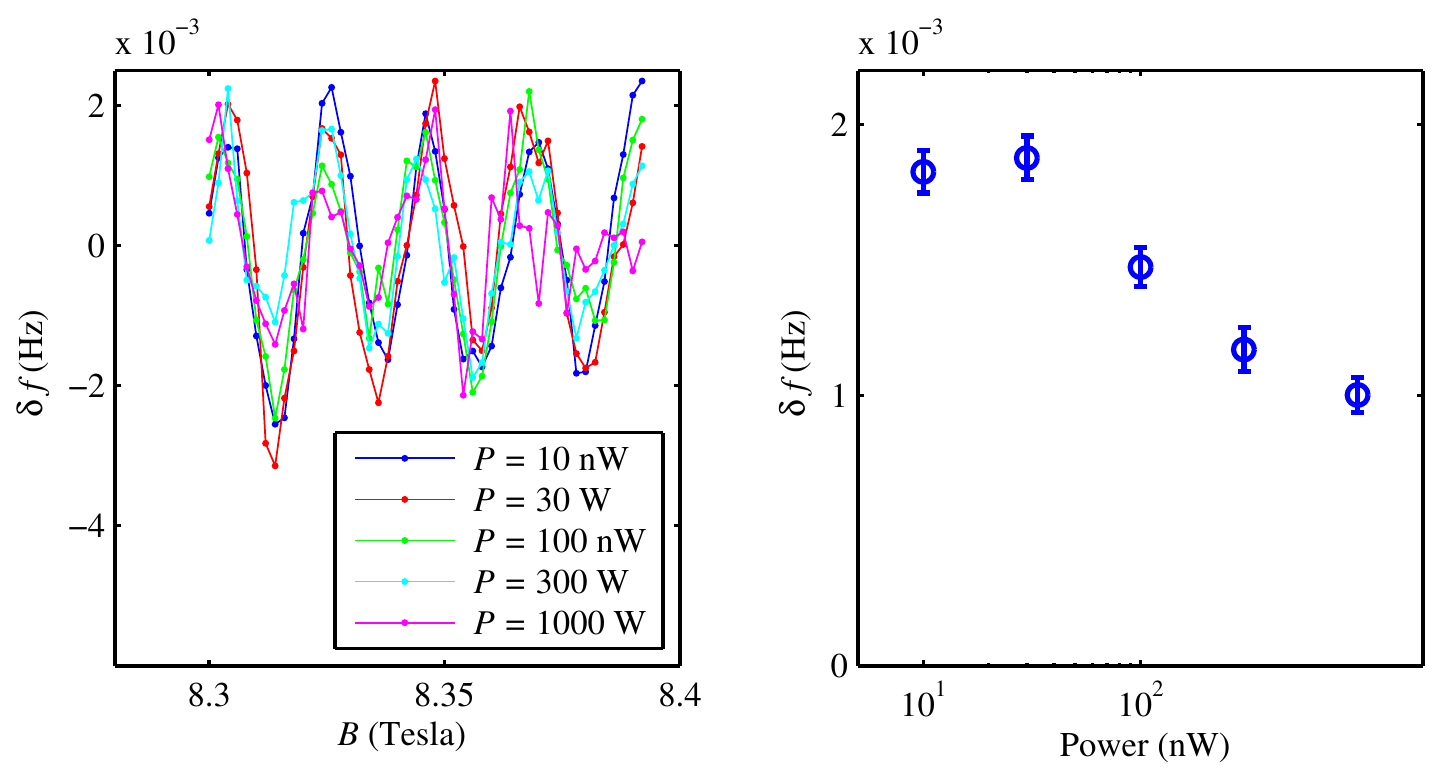}\caption{\label{fig:powerdep2}Left panel: Change in frequency versus magnetic
field for a series of laser powers incident on the cantilever. This
data was taken with sample 2. The drive was the same for all the traces.
Right panel: The mean amplitude of each of the traces from the left
panel, plotted versus laser power. For the data of samples 1, 2, 3
and 4 shown in the main paper, $10\text{ nW}$ of laser power was
used. }

\end{figure}
\begin{figure}
\begin{centering}
\includegraphics{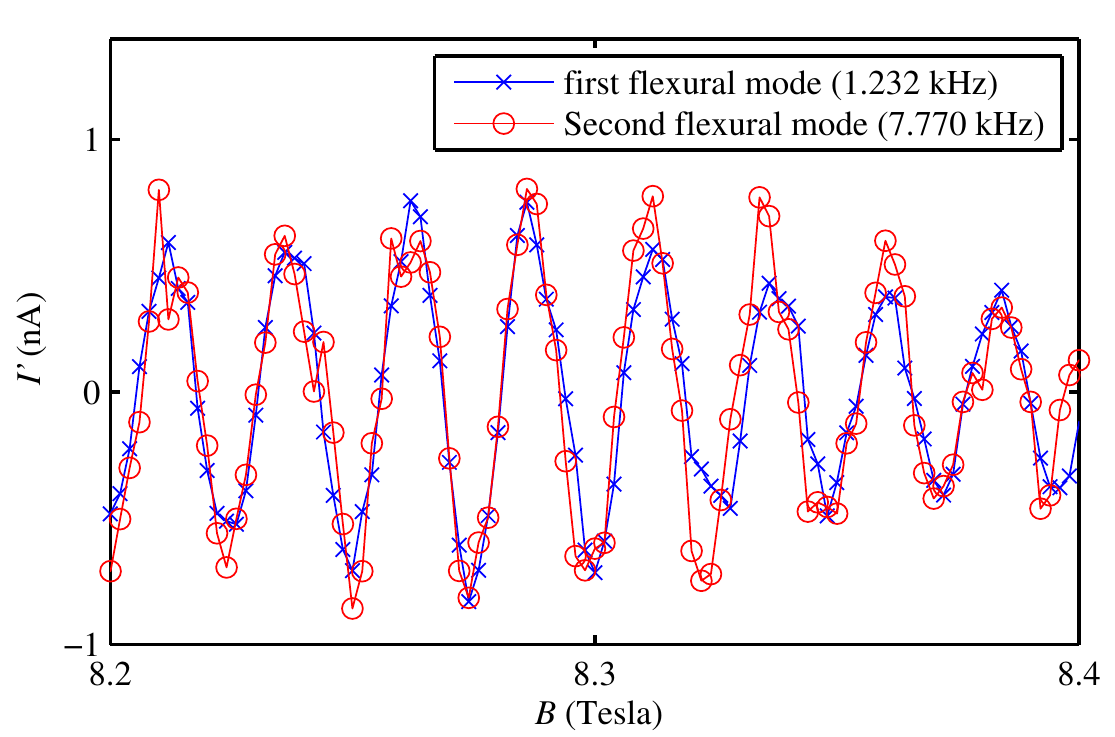}
\par\end{centering}

\begin{centering}
\caption{\label{fig:secondmode}The derivative of the persistent current I\textasciiacute{}
(derived from Eqs. \ref{eq:deltaf} and \ref{eq:Iprime}) versus magnetic
field measured when oscillating the cantilever at 1.232 kHz (the cantilever\textquoteright{}s
first flexural resonance) and at 7.77 kHz (the cantilever\textquoteright{}s
second flexural resonance). The persistent current does not appear
to depend on the cantilever oscillation frequency.}

\par\end{centering}

\end{figure}

\par\end{center}

\pagebreak{}

\section{Data of all the samples}

\subsection{Magnetic field sweeps}

\begin{figure}[H]
\begin{centering}
\includegraphics[clip]{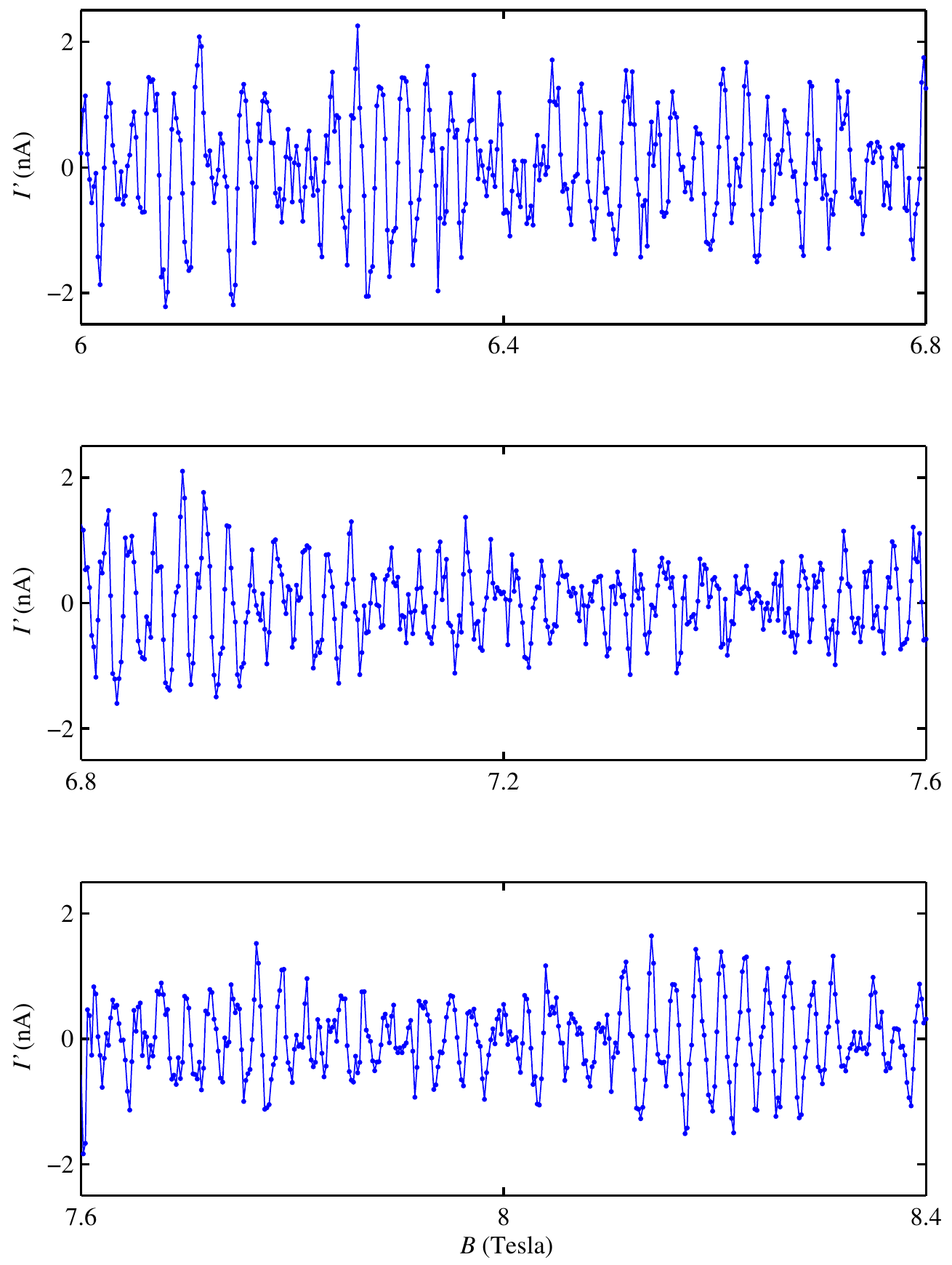}
\par\end{centering}

\centering{}\caption{\label{fig:CL04}The derivative of the persistent current $I'$ (derived
from Eqs. \ref{eq:deltaf} and \ref{eq:Iprime}) versus magnetic field
for sample 1 at T = 320 mK. The full sweep is separated into three
contiguous panels for clarity.}

\end{figure}

\begin{figure}
\begin{centering}
\includegraphics[clip]{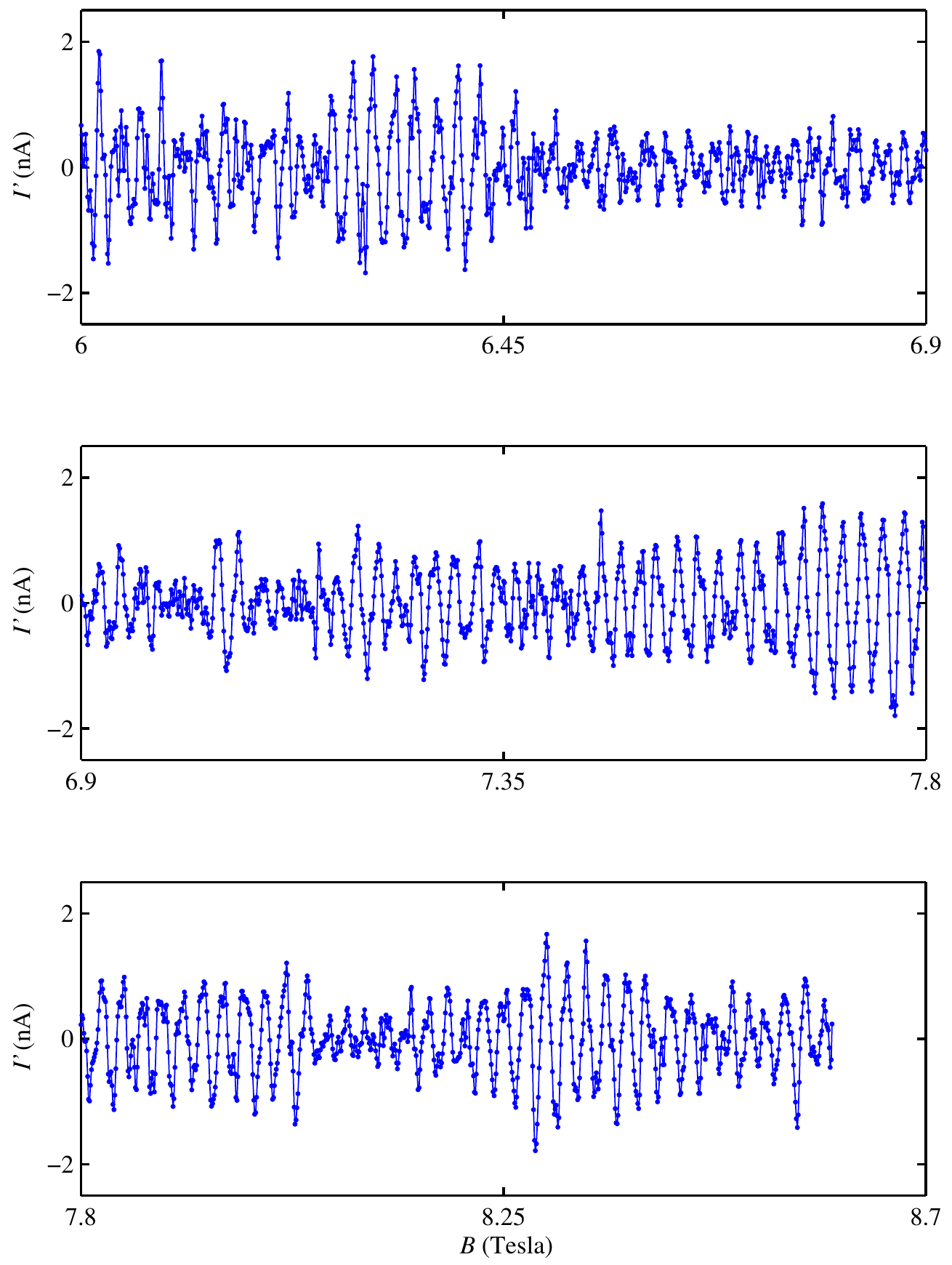}
\par\end{centering}

\centering{}\caption{\label{fig:CL06}The derivative of the persistent current $I'$ (derived
from Eqs. \ref{eq:deltaf} and \ref{eq:Iprime}) versus magnetic field
for sample 2 at T = 320 mK. }

\end{figure}

\begin{figure}[H]
\begin{centering}
\includegraphics[clip]{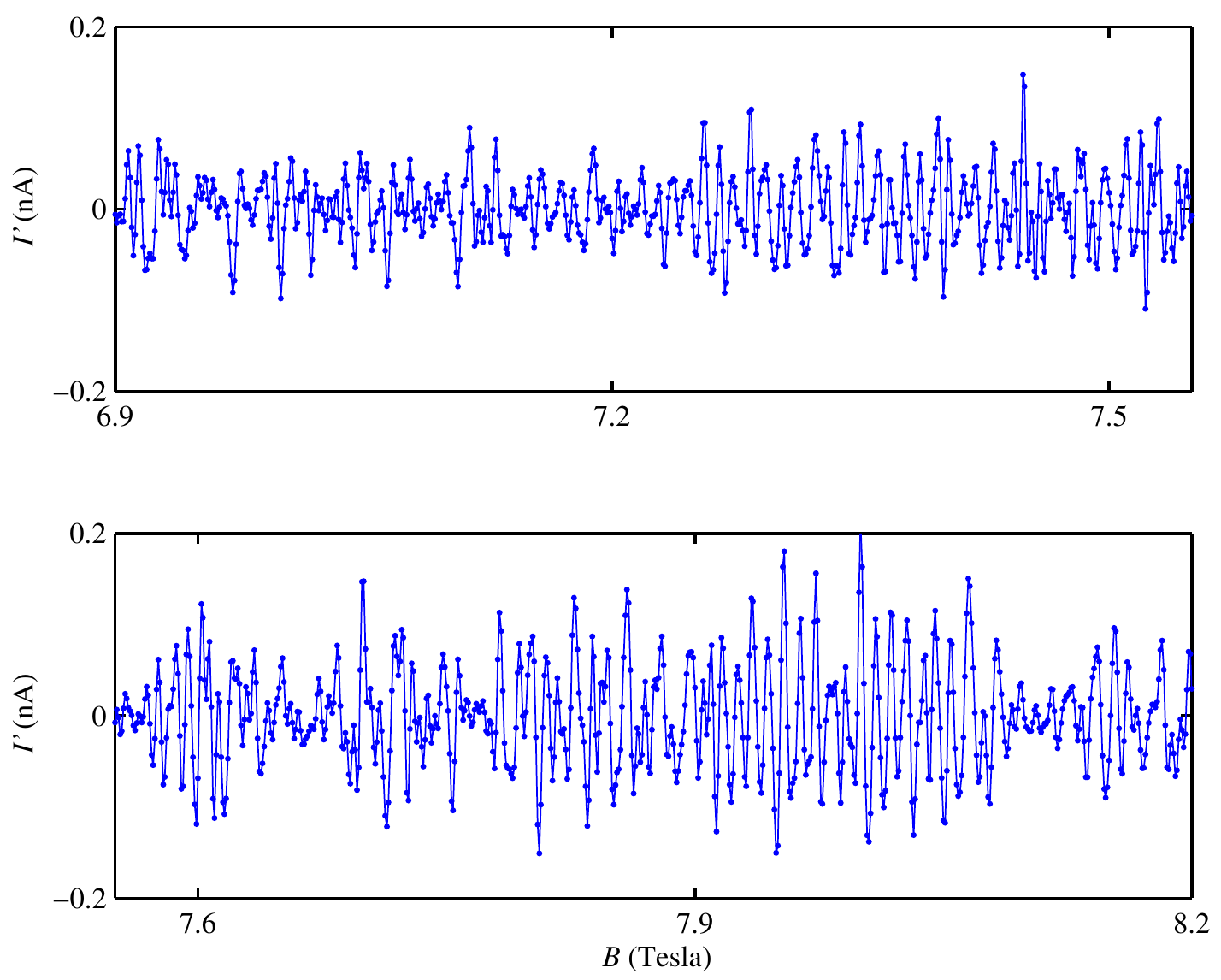}
\par\end{centering}

\centering{}\caption{\label{fig:CL31}The derivative of the persistent current $I'$ (derived
from Eqs. \ref{eq:deltaf} and \ref{eq:Iprime}) versus magnetic field
for sample 3 at T = 320 mK. }

\end{figure}

\begin{figure}
\begin{centering}
\includegraphics[clip]{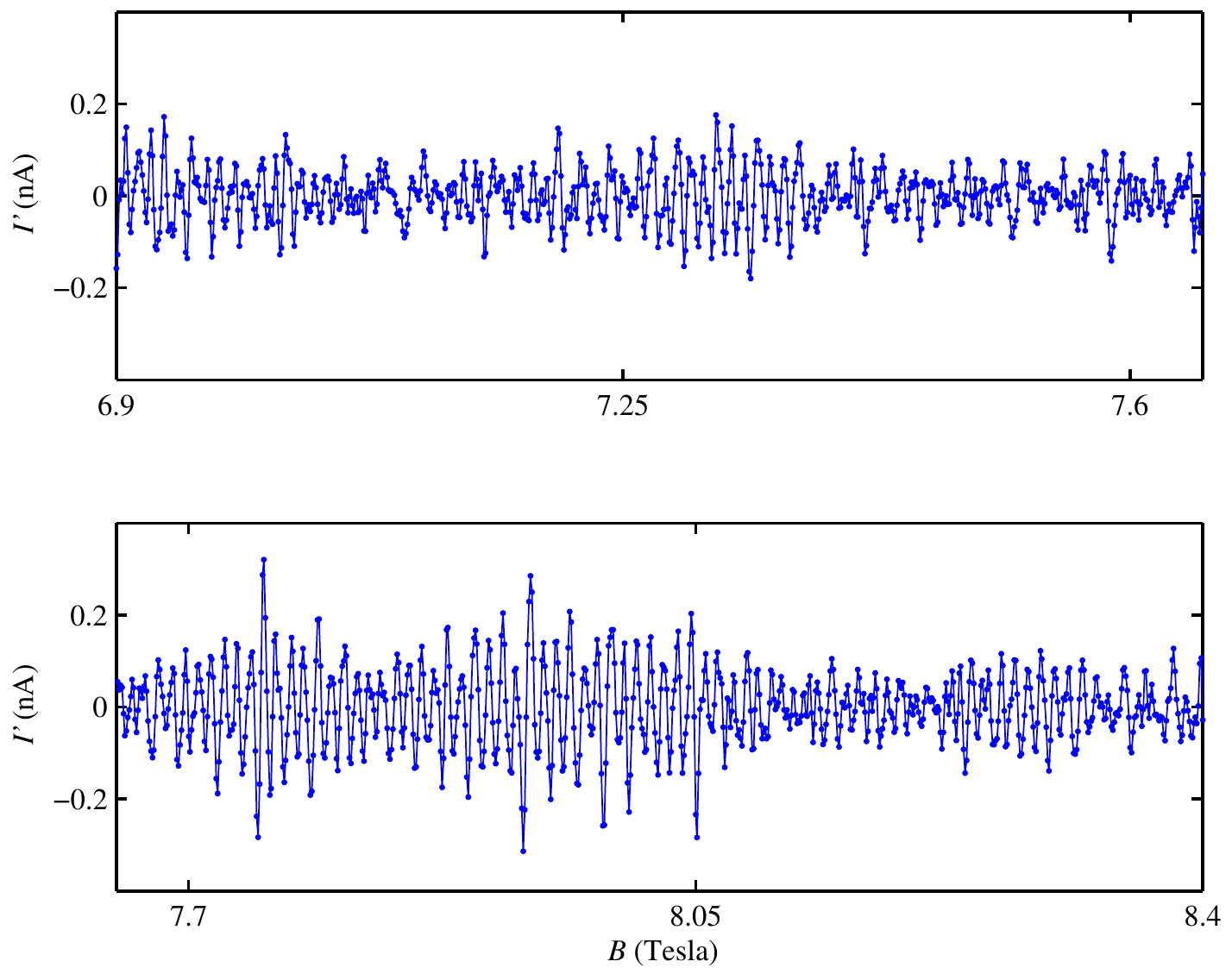}
\par\end{centering}

\centering{}\caption{\label{fig:CL32}The derivative of the persistent current $I'$ (derived
from Eqs. \ref{eq:deltaf} and \ref{eq:Iprime}) versus magnetic field
for sample 4 at T = 320 mK.}

\end{figure}

\begin{figure}[H]
\begin{centering}
\includegraphics[clip]{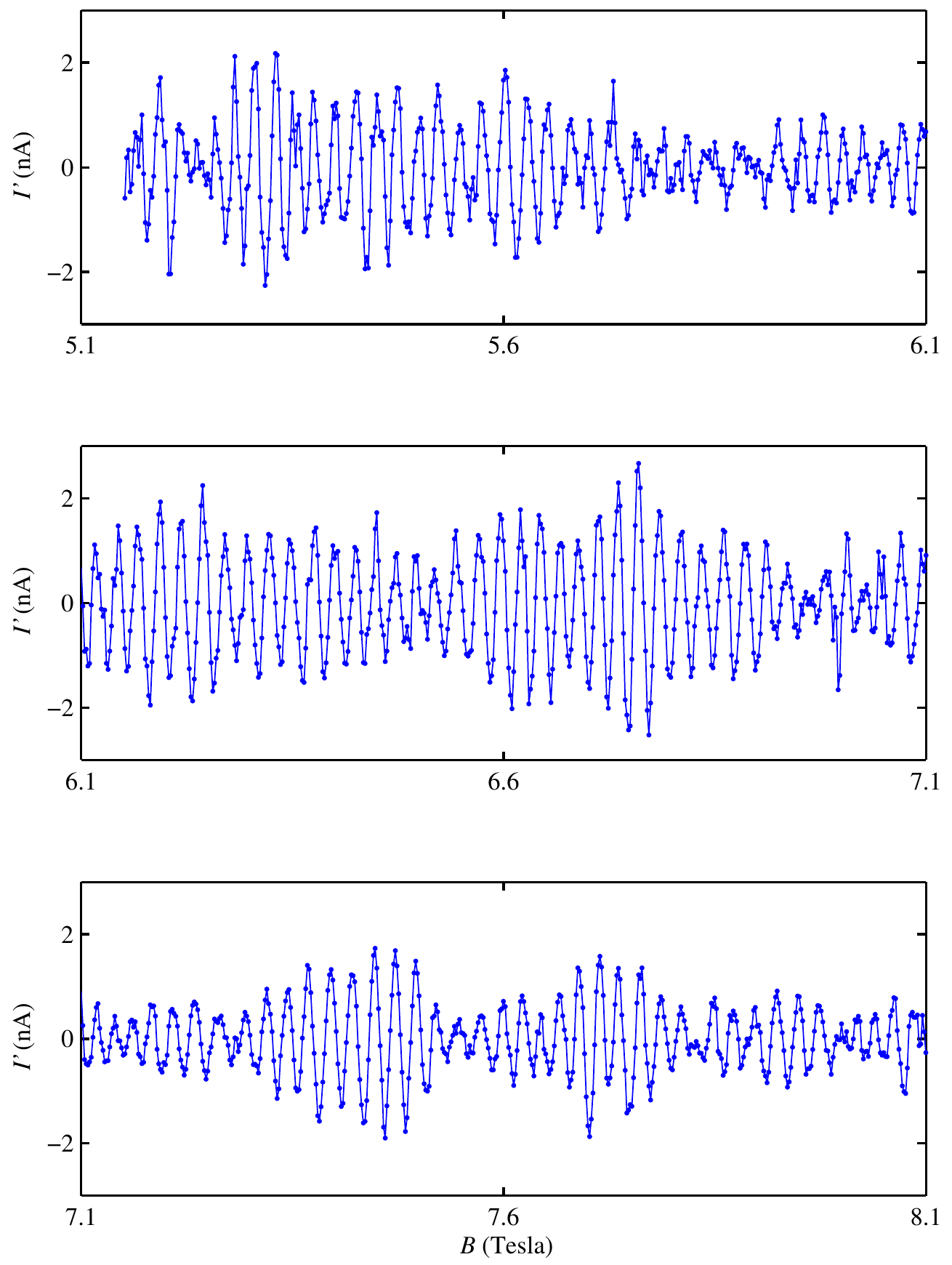}
\par\end{centering}

\centering{}\caption{\label{fig:CL61}The derivative of the persistent current $I'$ (derived
from Eqs. \ref{eq:deltaf} and \ref{eq:Iprime}) versus magnetic field
for sample 5 at T = 320 mK.}

\end{figure}

\begin{figure}[H]
\begin{centering}
\includegraphics[clip]{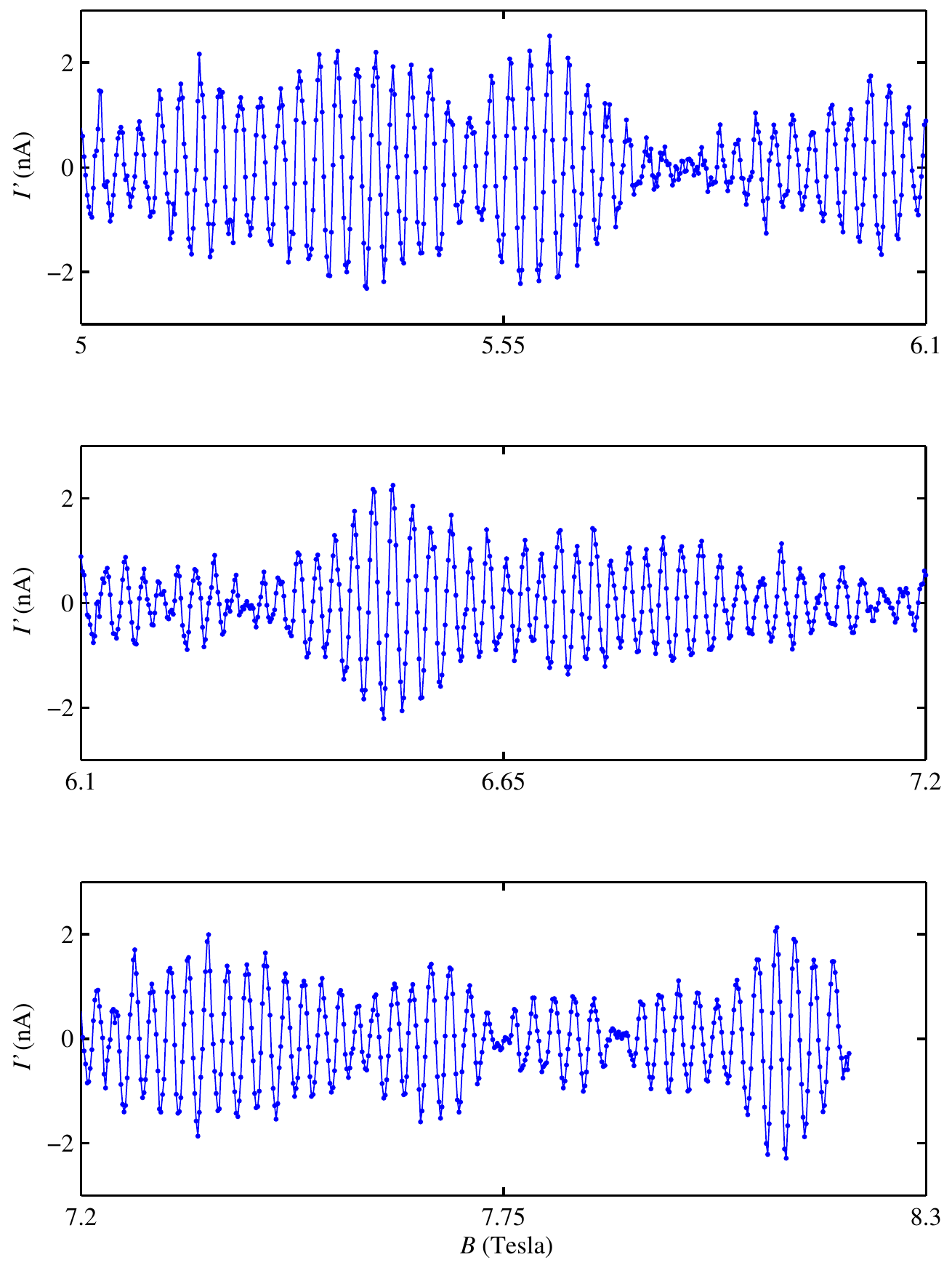}
\par\end{centering}

\centering{}\caption{\label{fig:CL62}The derivative of the persistent current $I'$ (derived
from Eqs. \ref{eq:deltaf} and \ref{eq:Iprime}) versus magnetic field
for sample 6 at T = 320 mK.}

\end{figure}

\begin{figure}
\begin{centering}
\includegraphics[clip]{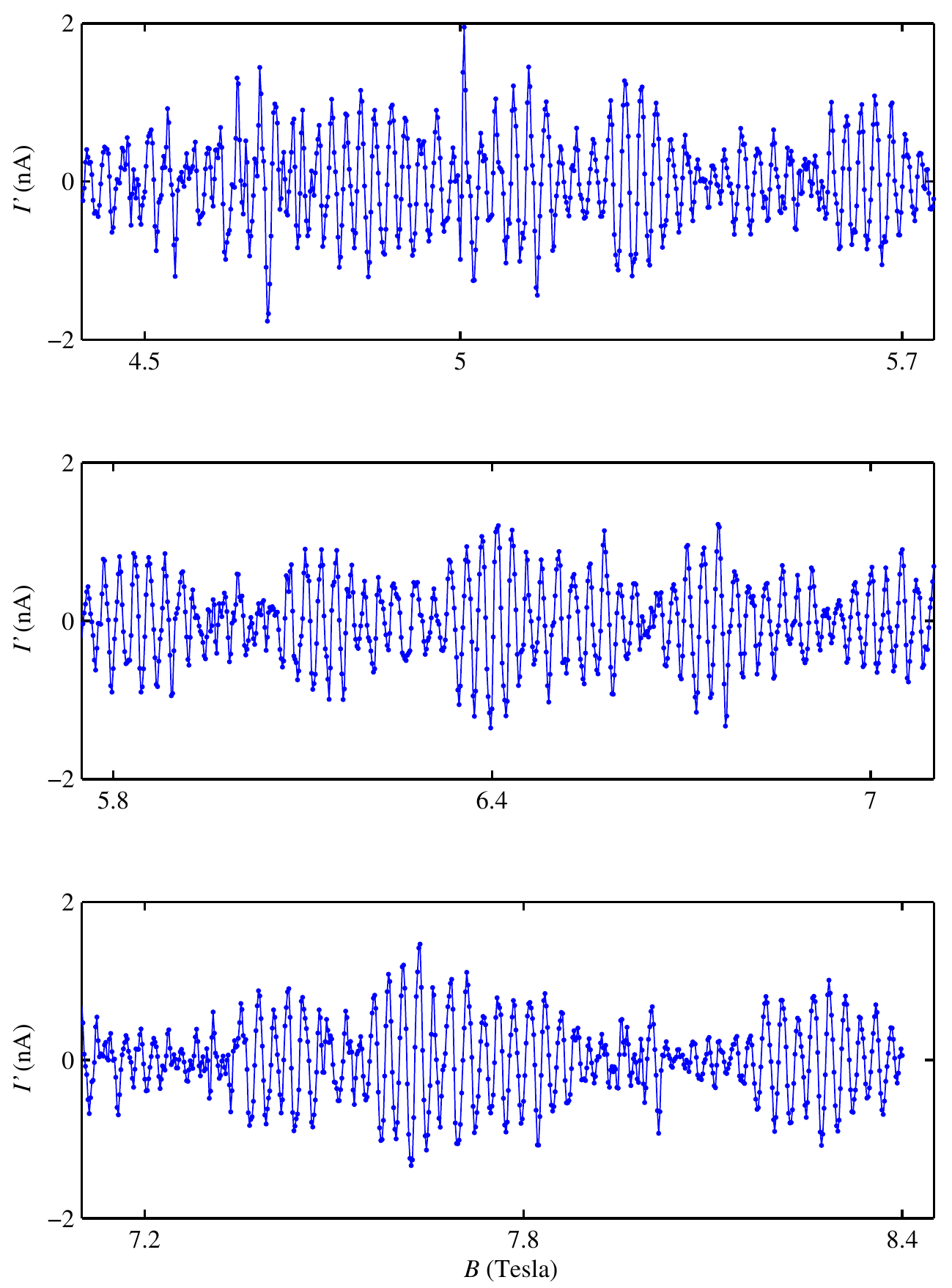}
\par\end{centering}

\centering{}\caption{\label{fig:CL63}The derivative of the persistent current $I'$ (derived
from Eqs. \ref{eq:deltaf} and \ref{eq:Iprime}) versus magnetic field
for sample 7 at T = 320 mK.}

\end{figure}

\begin{figure}[H]
\begin{centering}
\includegraphics[clip]{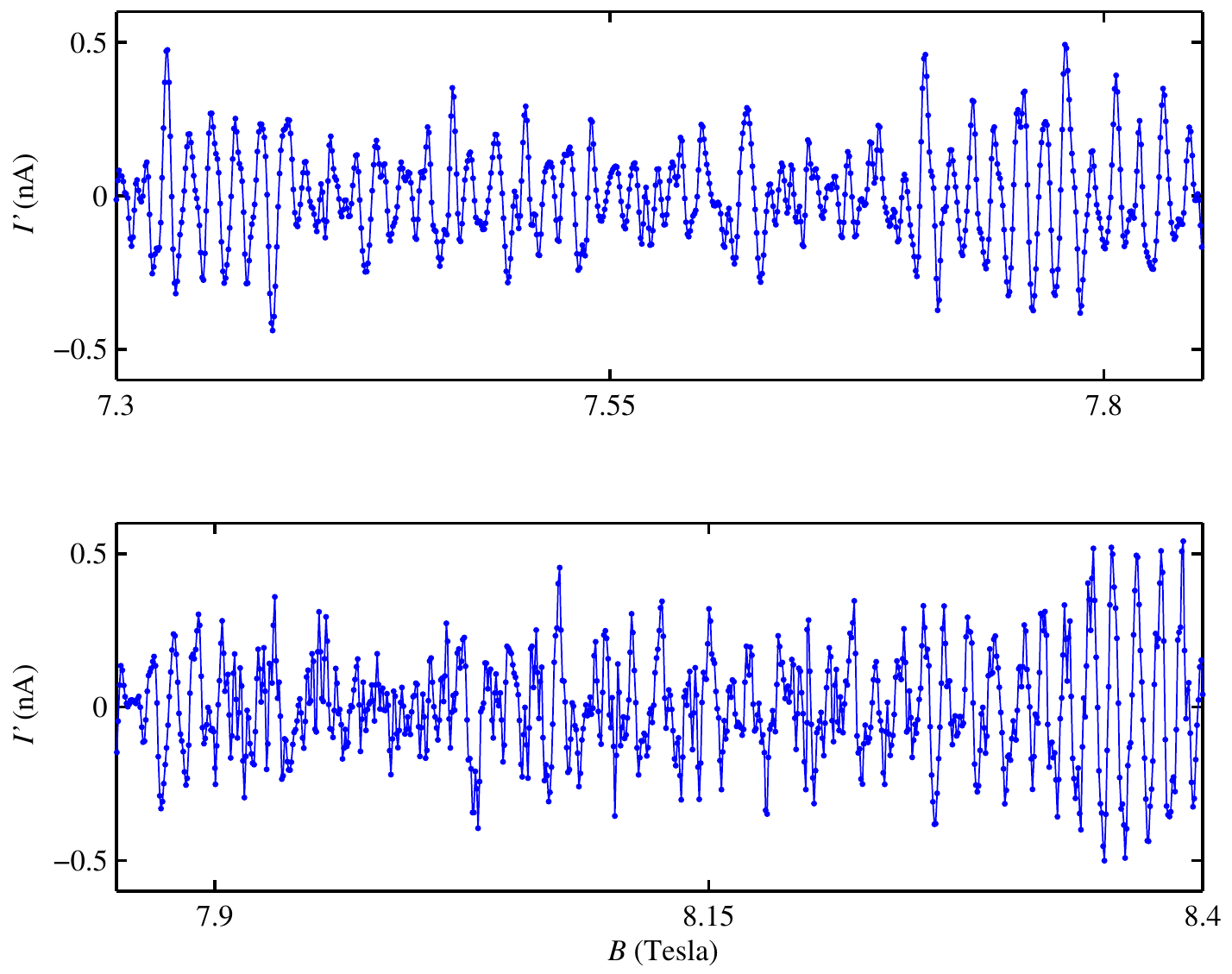}
\par\end{centering}

\centering{}\caption{\label{fig:CLWill}The derivative of the persistent current $I'$
(derived from Eqs. \ref{eq:deltaf} and \ref{eq:Iprime}) versus magnetic
field for sample 8 at T= 365 mK. }

\end{figure}

In this last section, we present figures \ref{fig:CL04}-\ref{fig:CLWill}
where we show the complete $I'$ versus magnetic field traces which
were analyzed in the main text. These traces were calculated, using
method B of Ref.\citet{Ania_Science} from measurements of the cantilever
frequency performed at the refrigerator's base temperature of 320
mK (365 mK for sample 8).

\pagebreak{}

\end{document}